\documentclass[twocolumn]{aa}
\usepackage[colorlinks,linkcolor={blue},citecolor={blue},urlcolor={red}]{hyperref}
\usepackage{graphicx}
\usepackage{natbib}
\usepackage{xspace}

\def\apjs{ApJS}

\def\beq{\begin{equation}}
\def\eeq{\end{equation}}
\def\bey{\begin{eqnarray}}
\def\eey{\end{eqnarray}}

\def\mpc{\, h^{-1}{\rm {Mpc}}}

\def\kpc{\, h^{-1}{\rm {kpc}}}

\def\kms{\,{\rm {km\, s^{-1}}}}

\def\msun{\, h^{-1}{\rm M_\odot}}
\def\msunt{\, h^{-2}{\rm M_\odot}}

\def\gs{\mathrel{\raise1.16pt\hbox{$>$}\kern-7.0pt
\lower3.06pt\hbox{{$\scriptstyle \sim$}}}}
\def\ls{\mathrel{\raise1.16pt\hbox{$<$}\kern-7.0pt
\lower3.06pt\hbox{{$\scriptstyle \sim$}}}}
\def\gtsima{\, {\buildrel > \over \sim} \,}
\def\ltsima{\, {\buildrel < \over \sim} \,}
\def\prosima{\, {\buildrel \propto \over \sim} \,}
\def\gsim{\lower.5ex\hbox{\gtsima}}
\def\lsim{\lower.5ex\hbox{\ltsima}}
\def\simgt{\lower.5ex\hbox{\gtsima}}
\def\simlt{\lower.5ex\hbox{\ltsima}}
\def\simpr{\lower.5ex\hbox{\prosima}}

\begin{document}
\title{Hosts and triggers of AGNs in the Local Universe}
\titlerunning{AGNs and Trigger}
\authorrunning{Zhang, Wang et al.}
\author{Ziwen Zhang\inst{1,2}, Huiyuan Wang\inst{1,2}, Wentao Luo\inst{3,1}, H.J. Mo\inst{4},   Zhixiong Liang\inst{1,2}, Ran Li\inst{5,6}, Xiaohu Yang\inst{7}, Tinggui Wang\inst{1,2}, Hongxin Zhang\inst{1,2}, Hui Hong\inst{1,2}, Xiaoyu Wang\inst{1,2}, Enci Wang\inst{8}, Pengfei Li\inst{1,2} \and JingJing Shi\inst{3}}
\institute{CAS Key Laboratory for Research in Galaxies and Cosmology, Department of Astronomy, University of Science and Technology of China, Hefei, Anhui 230026, China; Email: ziwen@mail.ustc.edu.cn, whywang@ustc.edu.cn
\and School of Astronomy and Space Science, University of Science and Technology of China, Hefei 230026, China
\and Institute for the Physics and Mathematics of the Universe (Kavli IPMU, WPI), UTIAS, Tokyo Institutes for Advanced Study, University of Tokyo, Chiba, 277-8583, Japan
\and Department of Astronomy, University of Massachusetts, Amherst MA 01003-9305, USA
\and Key laboratory for Computational Astrophysics, National Astronomical Observatories, Chinese Academy of Sciences, Beijing 100012, China
\and College of Astronomy and Space Sciences, University of Chinese Academy of Sciences, 19A Yuquan Road, Beijing 100049, China
\and Department of Astronomy, and Tsung-Dao Lee Institute, Shanghai Jiao Tong University, Shanghai 200240, China
\and Department of Physics, ETH Zurich, Wolfgang-Pauli-Strasse 27, CH-8093 Zurich, Switzerland}
\abstract{
Based on the spectroscopic and shear catalogs for SDSS galaxies in the local Universe, 
we compare optically-selected active galactic nuclei (AGNs) with control star-forming 
and quiescent galaxies on galactic, inter-halo and larger scales. 
We find that AGNs are preferentially found in two specific stages of 
galaxy evolution: star-burst and `green valley' phases, and that 
the stellar population of their host galaxies is quite independent of 
stellar mass, different from normal galaxies. 
Combining galaxy-galaxy lensing and galaxy clustering on large scales, 
we measure the mass of AGN host halos. 
The typical halo mass is about $10^{12}\msun$, similar to the
characteristic mass in the stellar mass-halo mass relation (SHMR). 
For given stellar mass, AGN host galaxies and star-forming galaxies 
share the same SHMR, while quiescent galaxies have more massive halos. 
Clustering analysis on halo scales reveals that AGNs are surrounded 
by a larger number of satellites (with stellar mass down to 1/1000 of the 
mass of the central galaxy) than star-forming galaxies, 
and that galaxies with larger stellar velocity dispersion have more satellites.
The number of satellites also increase with halo mass, reaching unity 
around $10^{12}\msun$. Our results suggest a scenario, in which the interaction 
of the central galaxy with the satellites triggers an 
early episode of star burst and AGN activities, followed 
by multiple AGN cycles driven by the non-axisymmetric structure
produced by the interaction. The feedback from the starburst 
and AGN reduces the amount of cold gas for fueling 
the central black hole, producing a characteristic 
halo mass scale, $\sim 10^{12}\msun$, where the AGN fraction peaks.
}

\maketitle

\section{Introduction}

In the local Universe, galaxies can be divided into 
two distinct populations: quiescent and star-forming 
galaxies \citep[e.g.][]{Strateva-01,Baldry-04,Brinchmann-04,Wetzel-Tinker-Conroy-12}. 
The number density of quiescent galaxies continuously 
increases with cosmic time since redshift of 
about four \citep[e.g.][]{Bell-04,Ilbert-13, Muzzin-13,Tomczak-14,Barro-17}, 
suggesting that galaxy quenching is an important process 
that drives galaxy evolution over most of the Hubble time.
To understand the underlying mechanisms, extensive studies 
have been carried out to search for the correlation of 
galaxy quenching with both internal properties of galaxies and their environments 
\citep[e.g.][]{Baldry-06,Weinmann-06,vandenBosch-08,Peng-10,Wetzel-Tinker-Conroy-12, 
Woo-13, Bluck-14, WangH-16, WangH-18, WangE-18a, Bluck-20, LiP20}. 

For central galaxies, which are the dominant galaxies in dark
matter halos, the most important internal and environmental 
parameters seem to be the central velocity dispersion of the galaxy
and the halo mass respectively \citep{Bluck-20}. 
This indicates that mechanisms that are related to the galaxy 
bulge mass or central black hole mass, such as the 
active galactic nuclei (AGN) feedback \citep[][]{Silk-Rees-98, Croton-06, Heckman-2014}, 
and those related to halo mass, such as galaxy 
interaction \citep[e.g.][]{Moore-96, Conselice-Chapman-Windhorst-03, DiMatteo-05} 
and virial shock heating \citep[e.g.][]{Dekel-Birnboim-06, Gabor-Dave-15}, 
may be responsible for the quenching. 
These mechanisms, in particular the AGN feedback, are expected to 
become dominant for massive galaxies, and help to yield 
a `pivot halo mass', $M_{\rm h, p}\sim10^{12}\msun$, in the stellar mass 
-halo mass relation (SHMR), at which, the efficiency for galaxy 
formation is maximum \citep[e.g.][]{Yang-Mo-vandenBosch-03, Wechsler-Tinker18}.

However the observational evidence for AGN feedback remains elusive. 
Indeed, AGN feedback has been reported to both enhance and suppress
star formation \citep[e.g.][]{Fabian-12, Mullaney-15, Delvecchio-15, Rodighiero-15, Kalfountzou-17, Mahoro-Povic-Nkundabakura-17, Bing19}.
One important reason for this uncertainty is that the lifetime of 
AGN activities, which is about $10^5$ to $10^8$ years 
\citep[e.g.][]{Marconi-04, Schawinski15, Yuan18}, 
is much shorter than the quenching time scale\citep[typically about 1 Gyrs,][]{Bell-04, Blanton06}, 
so that it is difficult to find an instantaneous correlation 
between AGN and star formation activities directly 
from observational data.

The locus of AGNs in the evolutionary path from star-forming 
galaxies to quiescent galaxies may provide valuable information 
about the role of AGNs. Previous studies have revealed some 
interesting trends. For example, some studies found that the host galaxies of 
AGNs appear to be located at the green valley, which is the 
transition region from star-forming to quiescent galaxies
\citep{Heckman-2014, Man19, Dodd2020}. Moreover, studies based on AGN 
clustering, weak lensing and galaxy groups all suggests that optically 
selected AGNs at low redshift reside preferentially in halos of roughly 
$M_{\rm h,A}=10^{12}\msun$ \citep[e.g.][]{Croom05, Pasquali-09, Mandelbaum09, Shen13}, 
similar to the pivot halo mass, suggesting  
that optical AGNs may be at a special stage of galaxy evolution.
It is thus interesting to understand why optical AGNs 
favor halos of $M_{\rm h,A}=10^{12}\msun$ and what processes, 
in these halos are responsible for triggering AGN 
activities. 

Many factors can affect the prevalence of AGN activities.
One important question is how to bring the gas down to 
the galaxy center to fuel the supermassive black holes (SMBH).
In the literature, two kinds of mechanisms are proposed.
One is the internal secular evolution process. The torque 
induced by non-axisymmetric galactic structures can 
drive slow and significant inflow \citep{Kormendy-04, Hopkins2011, Sellwood14, Fanali15}.
The galactic bar is one of the most prominent 
non-axisymmetric structures, and exists in about 40\% of 
spiral galaxies \citep{Oh-12}. And there is evidence
showing that bars can enhance star formation in the central 
regions of galaxies \citep[e.g.][]{Oh-12,Chown19}. However, 
whether galactic bars can significantly affect AGN 
activity is still under debate 
\citep{Arsenault89, Mulchaey-97, Oh-12, Galloway15, Goulding-17, Alonso18}.

Other mechanisms, such as galaxy merger and 
interaction, are also expected to displace the angular 
momentum of the gas and transport the gas inward
\citep[e.g.][]{Hopkins-06,DiMatteo-08,Bhowmick2020}.
Similar to the studies of secular evolution, observational 
evidence for this scenario is also mixed. 
Some studies found significant environmental dependence of AGN activities 
\citep[e.g.][]{Koulouridis-06,Koss-10,Ellison-11,Sabater-13,Khabiboulline-14,Lackner-14,Satyapal-14,Hong-15,Kocevski-15,Goulding-18,Gao-20}, 
while others found no or only weak environmental effects \citep[e.g.][]{Grogin-05,Li06b,Pierce-07,Ellison-08,Li-08,Gabor-09,Darg-10,WangL-19,Man19}.
The contradictory results 
may be caused by the difference in AGN selection criterion, 
observational bias, control sample and environmental indicator used. 
As we will show below, understanding environmental effects 
on AGNs also requires the knowledge about the evolutionary status of  
their host galaxies, as it can help us to better understand 
how to construct control samples and to adopt appropriate 
environmental indicators.

In this paper, we combine galaxy-galaxy weak lensing and galaxy 
clustering measurement to constrain the host halo masses of 
optically selected AGNs and their control samples. 
To take into account the galaxy evolution, we split the 
control sample into star-forming and quiescent galaxies.
We compare internal properties, small-scale clustering and halo 
mass of galaxies in the three samples to put AGNs in 
the evolutionary track of galaxy evolution and to 
understand the role of environmental processes. 

The paper is organized as follows. Section \ref{sec_ssm} presents
the AGN sample selection, control sample construction, and the 
methods of using galaxy clustering and galaxy-galaxy (g-g) lensing 
to derive halo mass. In Section \ref{sec_galpro},
we compare the properties of AGN host galaxies with those of control
samples. In Section \ref{sec_hmng}, we use g-g lensing and galaxy 
clustering to measure the mass of AGN host halos in comparison 
to that of the control samples. In Section \ref{sec_sat}, we analyze
satellites around AGNs and normal samples.  
We discuss the importance of using well-defined control samples, 
environmental triggering of AGN activities, and the connection of 
AGN feedback and galaxy evolution in Section \ref{sec_dis}. 
Finally, we summarize our results in Section \ref{sec_sum}.

\section{Samples and Methods of Analysis}\label{sec_ssm}

\subsection{AGN samples and control samples of normal galaxies}\label{sec_sample}

Our galaxy sample is drawn from the New York University Value Added 
Galaxy Catalog (NYU-VAGC) sample \citep{Blanton-05a}
of the Sloan Digital Sky Survey (SDSS) DR7 \citep{Abazajian-09}. In this paper, 
we mainly focus on central galaxies, which is defined as the most 
massive galaxies in galaxy groups. Here, we use the galaxy group 
catalog constructed by using the halo-based group-finding 
algorithm\citep{Yang-05, Yang-07} to separate centrals from satellites. 
Following \cite{Yang-07}, we select galaxies with r-band 
Petrosian magnitudes $r\le$ 17.72,  with redshifts in the range 
$0.01 \le z \le 0.2$, and with redshift 
completeness $C_{z} >$ 0.7. Stellar mass of individual 
galaxies, $M_*$, are obtained using the relation between 
the stellar mass-to-light ratio and color, as given by \citet{Bell-03}, 
but assuming a Kroupa IMF \citep{kroupa-Pavel-01}.  
This leads to a $-0.1$ dex correction in the stellar mass-to-light 
ratios relative to the original values. To obtain the 
star formation rate (SFR) and 4000\AA~break ($D_n4000$) of 
individual galaxies, we combine our galaxy sample with the 
MPA/JHU SDSS catalog\citep{Brinchmann-04}. 
The total galaxy sample (tG) contains 593,227 galaxies, of which
452,177 are identified as centrals (hereafter cG).

Active galactic nuclei (AGNs) are identified using the BPT 
diagram \citep{Baldwin-Phillips-Terlevich-81} from the tG sample. 
In particular, we use the demarcation line proposed by \cite{Kauffmann-AGN-03},
in the $\rm [OIII]\lambda5007/H\beta$ versus $\rm [NII]\lambda6583/H\alpha$
diagram. The fluxes of the four emission lines are taken from the 
MPA/JHU catalog. Following \cite{Brinchmann-04}, we require the four 
spectral lines to have a signal-to-noise ratio greater than 3.0. 
These selection criteria result in a total of 57,252 AGNs 
(hereafter tAGN sample). Among them, 46,198 are central galaxies, 
and the corresponding sample is denoted by cAGN. 

A control sample of galaxies is constructed by simultaneously 
matching both the redshift and the stellar mass ($M_*$). 
The adopted tolerances in the matching are $|\Delta z|< 0.005$
and $|\Delta\log_{10}M_{*}|< 0.1$. For each AGN, four control galaxies are 
selected. Several types of control samples are constructed. 
For the tAGN sample, we construct a control sample, tG$^{\rm c}$, 
from the tG sample. For the cAGN sample, a control sample, cG$^{\rm c}$, 
is constructed from the cG sample. We also separate galaxies into 
a star-forming population and a quiescent population in the SFR-$M_*$ space, 
using the demarcation line proposed by \cite{Bluck-16}. 
Thus, for the cAGN sample, we also construct a control
star-forming (cSF$^{\rm c}$) sample and a control quiescent (cQ$^{\rm c}$) sample.

As shown below, the stellar velocity dispersion ($\sigma_*$) for AGNs is 
systematically different from other galaxies of the same stellar mass.
We thus also construct control samples according to $\sigma_*$.
The values of $\sigma_*$ are also taken from the NYU-VAGC and corrected 
to the same effective aperture using the formula of \cite{Cappellari-06}.
The control samples, cSF$^{\rm c2}_{\sigma_*}$ and cQ$^{\rm c2}_{\sigma_*}$, 
are constructed, respectively, from the star-forming and quiescent 
galaxies with $\sigma_*$ measurements, by matching redshift, $M_*$ and 
$\sigma_*$. The tolerance in $\sigma_*$ is $\mid\Delta\sigma_{*}\mid<$ 20$\kms$.
Some AGNs have no $\sigma_*$ measurements and/or no matched galaxies. 
Excluding these galaxies results in 43,851 central AGNs, and this 
new AGN sample is referred to as cAGN$_{\sigma_*}$. For comparison,
we also construct another set of control samples, cSF$^{\rm c}_{\sigma_*}$ 
and cQ$^{\rm c}_{\sigma_*}$, for cAGN$_{\sigma_*}$, by only matching 
stellar mass and redshift.

The first lowercase letter, `t' or `c', in the sample name 
indicates that the sample includes both centrals and satellites (total)
or only centrals. The superscript, `c', indicates the control samples with stellar
mass and redshift controlled, while `c2' indicates the control samples with
$\sigma_*$ additionally controlled. If a sample has no superscript, it is not
a control sample, such as cAGN and cG. Most of our following analyses focus 
on the cAGN sample and its control samples.

\subsection{The cross-correlation analysis}
\label{sec_ccf}

The auto-correlation functions of AGNs and the 
AGN-galaxy cross correlation function provide effective ways to study the 
large scale environments of AGNs \citep[e.g.][]{Croom05,Li06b, Shen13, Zhang2013,
Jiang2016, Laurent17, Shankar2019}. Here we use the projected two-point 
cross-correlation function (hereafter 2PCCF) to quantify the clustering of 
our selected sample with respect to the corresponding reference galaxy sample. 
The reference sample is constructed in exactly the same way 
as described in \cite{WangL-19}, and here we provide a brief description
about the construction. The reference sample is a 
magnitude-limited sample selected from the NYU-VAGC sample\citep{Blanton-05a}. 
It consists of 510,605 galaxies with $r$-band Petrosian
apparent magnitude of $r<17.6$, with $-24 < M_{0.1_{r}} < -16$, 
and with spectroscopic redshift in range $0.01 < z < 0.2$. 
Here, $M_{0.1_{r}}$ is the $r$-band Petrosian absolute magnitude, 
$K+E$-corrected to $z=0.1$.
The random sample is constructed following the method described in 
\cite{Li06a}. For each galaxy in the reference sample, we
duplicate it at 10 randomly-selected sky positions in the SDSS survey 
area,\footnote{The geometry of the survey area is described by a set 
of spherical polygons, see
{\tt http://sdss.physics.nyu.edu/vagc/} \citep{Blanton-05a}.}
keeping all other properties (including redshift) of the galaxy 
unchanged. The resulted random sample has the same survey geometry,
the same distribution of galaxy intrinsic properties, and the 
same redshift distribution as the reference sample. 
The 2PCCF is statistically more robust than the auto-correlation 
function,  because we can use the large number of reference galaxies
to determine both the small and large scale environments of AGNs. 
The 2PCCF on small scales describes the abundance of 
neighboring galaxies around the selected galaxies, 
and that on large scales carries information about halo bias, 
thereby providing constraints on the host halo mass of 
AGNs \citep[e.g.][]{Mo-White-96}. 

We estimate the 2PCCF,  $\xi (r_{\rm p},\pi)$,  using
\begin{align}
\xi (r_{\rm p},\pi) = \frac{N_{\rm R}}{N_{\rm D}}\frac{GD(r_{\rm p},\pi)}{GR(r_{\rm p},\pi)}-1,
\end{align}
where $N_{\rm D}$ and $N_{\rm R}$ are the galaxy numbers in the 
reference and random samples, respectively; $r_{\rm p}$ and $\pi$ 
are the separations perpendicular and parallel to the line 
of sight, respectively; $GD$ is the number of cross pairs between 
the selected  sample and the reference sample; $GR$ is that 
between the selected sample and the random sample. 

Integrating $\xi (r_{\rm p},\pi)$ along the line of sight to
reduce the redshift distortion effect, we  obtain the projected 2PCCF, 
\begin{align}
    { w_{\rm p}(r_{\rm p}}) 
    = 2\int_{0}^{\infty}\xi(r_{\rm p},\pi)d\pi 
    = 2\sum_{i}\xi(r_{\rm p},\pi_{i})\Delta \pi_{i},
\end{align}
where $\pi_{i}$ and $\Delta \pi_{i}$ are the
separation parallel to the line of sight and the corresponding bin size. 
We adopt $\pi_{\rm max}$ = 40 $\mpc$ as the 
upper limit of the integration and $\Delta \pi_{i}$ = 1 $\mpc$.
We sample $r_{\rm p}$ in 10 logarithmic bins with 
$r_{\rm p,min} = 0.01 \mpc$ and $\Delta\log (r_{\rm p}/\mpc) = 0.345$.
The errors on the measurements of the 2PCCF are estimated by using 100 
bootstrap samples \citep{Barrow-Bhavsar-Sonoda-84}. We correct the 
fiber collision effects by using the same method as in \cite{Li06a}, and we 
refer the reader to the original paper for details and validity tests.

\subsection{Weak-lensing shear measurements and halo-mass estimates}
\label{sec_gg}

The shear catalog used here is created by \cite{Luo17}. Their selection of 
source galaxies is from SDSS DR7 image data in the {\itshape r} band, 
which covers about 8423 square degrees of the SDSS LEGACY sky.
A sequence of Flags and model magnitude cuts with {\itshape r} $\le 22.0$ and 
{\itshape i} $\le 21.5$ are applied to the image data.  
The shapes of the galaxy images are obtained, and the final shape catalog
consists of the shape measurements with the resolution factor {\itshape R} 
equal or greater than $1/3$.  This shape catalog contains 39,625,244 
galaxies with positions, shapes, and photo-$z$ information for individual 
source galaxies.

We measure the galaxy-galaxy lensing signal by stacking the tangential ellipticity 
of source galaxies in projected radial bins 
\citep{Miralda-Escude-91, Sheldon-04, Mandelbaum-05, Mandelbaum09, Luo-18}.
The non-zero tangential ellipticity, a.k.a the tangential shear ($\gamma_{\rm t}$) 
is related to the excess surface density (ESD), $\Delta \Sigma$, by
\begin{align}
    &\Delta \Sigma(r_{\rm p}) = \gamma_{\rm t}\Sigma_{\rm crit} = \Bar{\Sigma}(< r_{\rm p}) - \Sigma(r_{\rm p}),
\end{align}
where $\Bar{\Sigma}(< r_{\rm p})$ is the average surface mass density within 
$r_{\rm p}$, $\Sigma(r_{\rm p})$ is the surface mass density at $r_{\rm p}$, and 
$\Sigma_{\rm crit}$ is the geometrical factor defined as
\begin{align}
    &\Sigma_{\rm crit} = \frac{c^{2}}{4\pi G} \frac{D_{\rm s}}{D_{\rm l}D_{\rm ls}(1+z_{\rm l})^2},
\end{align}
where $c$ is the speed of light, $G$ is the gravity constant, $z_{\rm l}$ is the redshift 
of the lens, $D_{\rm ls}$ is the angular diameter distance between the lens and the source, 
$D_{\rm l}$ and $D_{\rm s}$ are the angular diameter distances of the lens and the source, 
respectively. In addition, we estimate the errors of the lensing signal by using 2,500 bootstrap samples.

We use two models to fit the weak lensing signal around galaxies.
The first one (hereafter M1) assumes that the lensing signal as the 
combination of three terms,
\begin{align}
    &\Delta \Sigma(r_{\rm p}) = \Delta \Sigma_{\rm NFW}^{\rm off}(r_{\rm p}) 
    + \frac{M_{*}}{\pi r_{\rm p}^{2}} 
    + \Delta \Sigma_{\rm 2h},
\end{align}
where the first term is the one halo term taking into account the possibility 
that central galaxies may not be located at the centers of their host halos, 
the second term is the contribution from the stellar mass of the central galaxy, 
and the third term is the projected two-halo term. Since we only apply the
method to central galaxies, cAGN and cAGN$_{\sigma_*}$ and their control 
galaxies, we do not include the satellite component.

\cite{Yang-06} provided the analytical formulae to calculate the ESD of 
the one-halo term from a NFW profile \citep{Navarro-Frenk-White1997}
that is specified by two free parameters, the halo mass $M_{\rm h}$ 
and the concentration. We adopt their formula for the ESD with an additional 
parameter, $R_{\rm off}$, that specifies the projected off-center distance.
Following the model proposed by \cite{Johnston-07},
we describe $R_{\rm off}$ by a two-dimensional Gaussian 
distribution with mean equal to zero and dispersion given by $\sigma_{\rm off}$. 
To model the two-halo term, we first use  
CAMB\footnote{https://camb.info/} (Code for Anisotropies from 
Microwave Background) \citep{Lewis-11} 
and the mcfit\footnote{https://github.com/eelregit/mcfit/} 
package \citep{LiYin-19} to obtain the matter correlation 
function, $\xi_{\rm mm}(r)$. We then use the halo bias model of 
\citet{Tinker-10} to obtain the bias factor,  
$b_{\rm h}(M_{\rm h})$, and to calculate the halo-matter cross-correlation 
function, $\xi_{\rm hm}(r)=b_{\rm h}\xi_{\rm mm}(r)$.
The projected two-halo term is obtained directly from 
$\xi_{\rm hm}$ \citep{Cacciato-09}. Finally, the stellar component 
is modelled as a point source and the stellar mass parameter is fixed 
as the mean value of $M_*$ of the galaxy sample. 
We refer the reader to \cite{Luo-18} for a detailed description 
about the modelling of the three components.

Thus, model M1 consists of three free parameters, halo mass $M_{\rm h}$, 
halo concentration and $\sigma_{\rm off}$. We use 
emcee\footnote{https://emcee.readthedocs.io/en/stable/} \citep{Foreman-13} to 
run a Monte Carlo Markov Chain (hereafter MCMC) to constrain these parameters, 
assuming the following likelihood function,
\begin{align}
    &\ln(\mathcal{L}1) = -{1\over 2} (\Delta \Sigma_{\rm l}-\Delta \Sigma_{\rm m})^{T}C_{1}^{-1}(\Delta \Sigma_{\rm l}-\Delta \Sigma_{\rm m}),\label{eq_llLense}
\end{align}
where $\Delta \Sigma_{\rm l}$ and $\Delta \Sigma_{\rm m}$ represent the true 
lensing signal and the model, respectively, and $C_{1}^{-1}$ is the inverse 
of the covariance matrix.  We only use the trace components of the covariance 
matrix to construct the likelihood function for the following two reasons.
First, at scales smaller than our ESD measurements, shape noise dominates the 
error budget.  Second, the covariance matrix is too noisy to be modeled 
reliably \citep{Viola-15}. The priors of the three parameters are set to be 
flat,  with the halo mass in the range [11.0, 16.0] in logarithmic space, 
the concentration in the range [1.0, 16.0], and $\sigma_{\rm off}$ in the 
range of [0.001, 0.3] in units of the virial radius. 
In running the emcee, we use 300 walkers and run a chain of 5000 steps 
with 500 burn-in steps, starting from an initial setting of the three parameters, 
$\log(M_{\rm h}/\msun)=12.8$, concentration $=7.9$, and $\sigma_{\rm off}=0.09$.

For the second model (hereafter M2), we combine the results from 
weak lensing and 2PCCF to constrain the halo mass. 
Different from M1, here we use the MCMC to fit the lensing results of 
cAGN, cSF$^{\rm c}$ and cQ$^{\rm c}$ simultaneously, and use the 
ratios of the 2PCCFs at large scales between the three samples 
as additional constraints. To this end, we use the halo mass 
estimated at each MCMC chain step to calculate the halo bias from 
the analytical formula given in \cite{Tinker-10}. We then obtain 
the model bias ratios, cAGN$/$cSF$^{\rm c}$ and cAGN$/$cQ$^{\rm c}$, 
and fit them to the corresponding ratios obtained from the observed 2PCCF. 
The likelihood function for the bias term is similar to 
Equation (\ref{eq_llLense}), except that the
covariance matrix $C_{2}$ is built from bootstrap sampling,
\begin{align}
    &\ln(\mathcal{R}) = -{1\over 2}(R_{w_{\rm p}}-R_{\rm hb})^{T}C_{2}^{-1}(R_{w_{\rm p}}-R_{\rm hb}),
    \label{eq_ratio}
\end{align}
where $R_{\rm hb}$ and $R_{w_{\rm p}}$ are the model bias ratio 
and the 2PCCF ratio between AGNs and the corresponding 
control sample, respectively. We only use the 2PCCF ratios 
on large-scales in the fitting: $r_{\rm p}>1\mpc$ for cAGN$/$cSF$^{\rm c}$ 
and $r_{\rm p}>4\mpc$ for cAGN$/$cQ$^{\rm c}$. 
The reason for these choices and the robustness of the method are 
described in Section \ref{sec_mh2PCCF}. 

 Model M2 is thus described by five likelihood terms in each MCMC step, 
three from the weak-lensing constraints and two from the 2PCCF ratios:
\begin{align}
    \begin{split}
    &\ln(\mathcal{L}2) \\
    &=\ln(\mathcal{L}1)_{\rm cAGN} + \ln(\mathcal{L}1)_{\rm cSF^{\rm c}} + \ln(\mathcal{L}1)_{\rm cQ^{\rm c}} \\
    &+\ln(\mathcal{R})_{\rm cAGN/cSF^{\rm c}} + \ln(\mathcal{R})_{\rm cAGN/cQ^{\rm c}},
    \end{split}
\end{align}
The value of $\ln(\mathcal{L}2)$ at a given step is returned to the MCMC to decide the next 
chain step. The priors of $M_{\rm h}$, concentration and 
$\sigma_{\rm off}$ and the initial settings of the MCMC for M2 are the same 
as for M1.

\section{Properties of AGN host galaxies in comparison 
to normal star-forming and quiescent galaxies}
\label{sec_galpro}

\begin{figure*}
    \centering
    \includegraphics[scale=0.9]{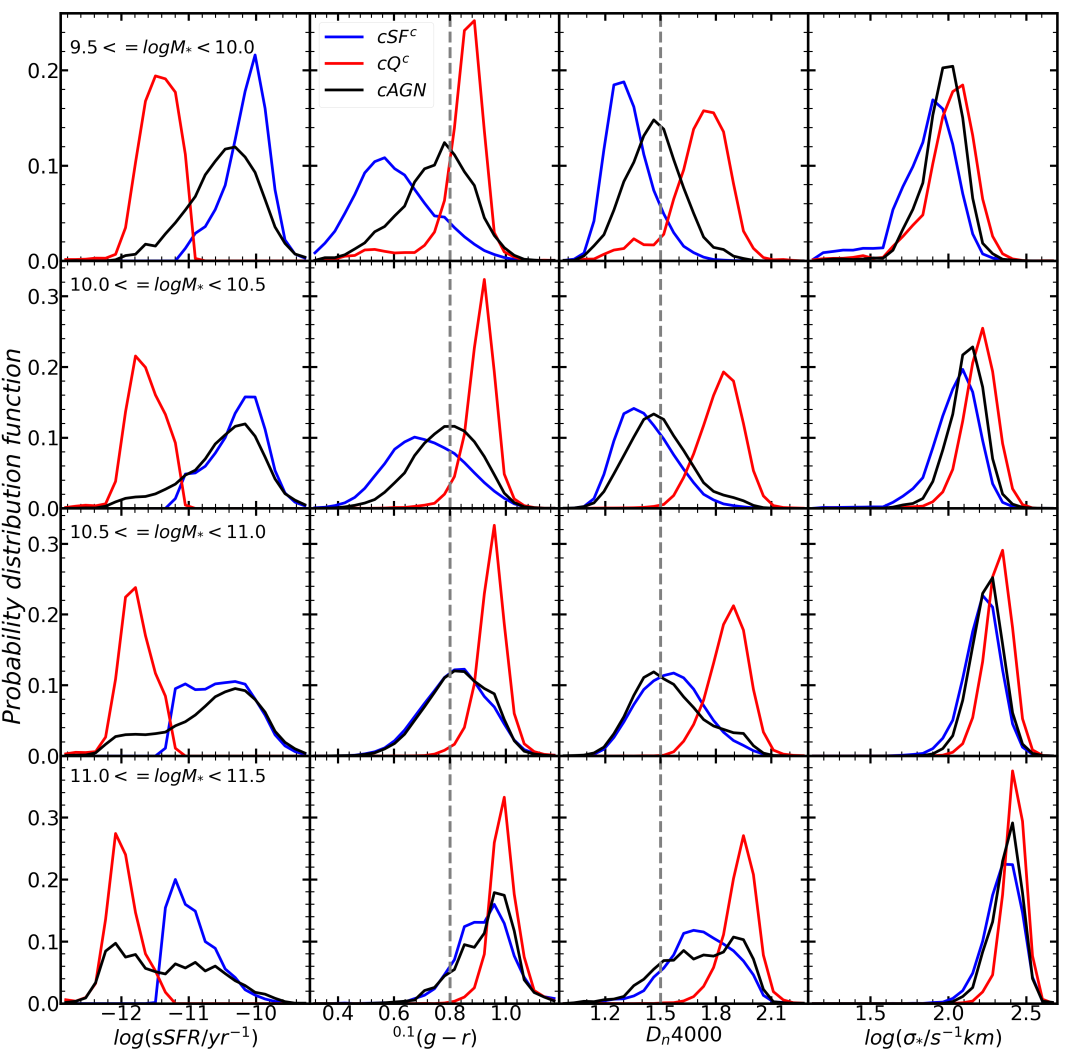}
    \caption{The probability distribution functions in different stellar mass bins (as indicated in each row) of sSFR, color, $D_{n}$4000 and central velocity dispersion (each column) for cAGN (black), cSF$^{\rm c}$ (blue) and cQ$^{\rm c}$ (red). In the middle two columns, the vertical dashed lines (grey) show $^{0.1}(g-r)$=0.8 (second column) and $D_{n}$4000=1.5 (third column). }\label{fig_galpro}
\end{figure*}

\begin{figure}
    \centering
    \includegraphics[scale=0.3]{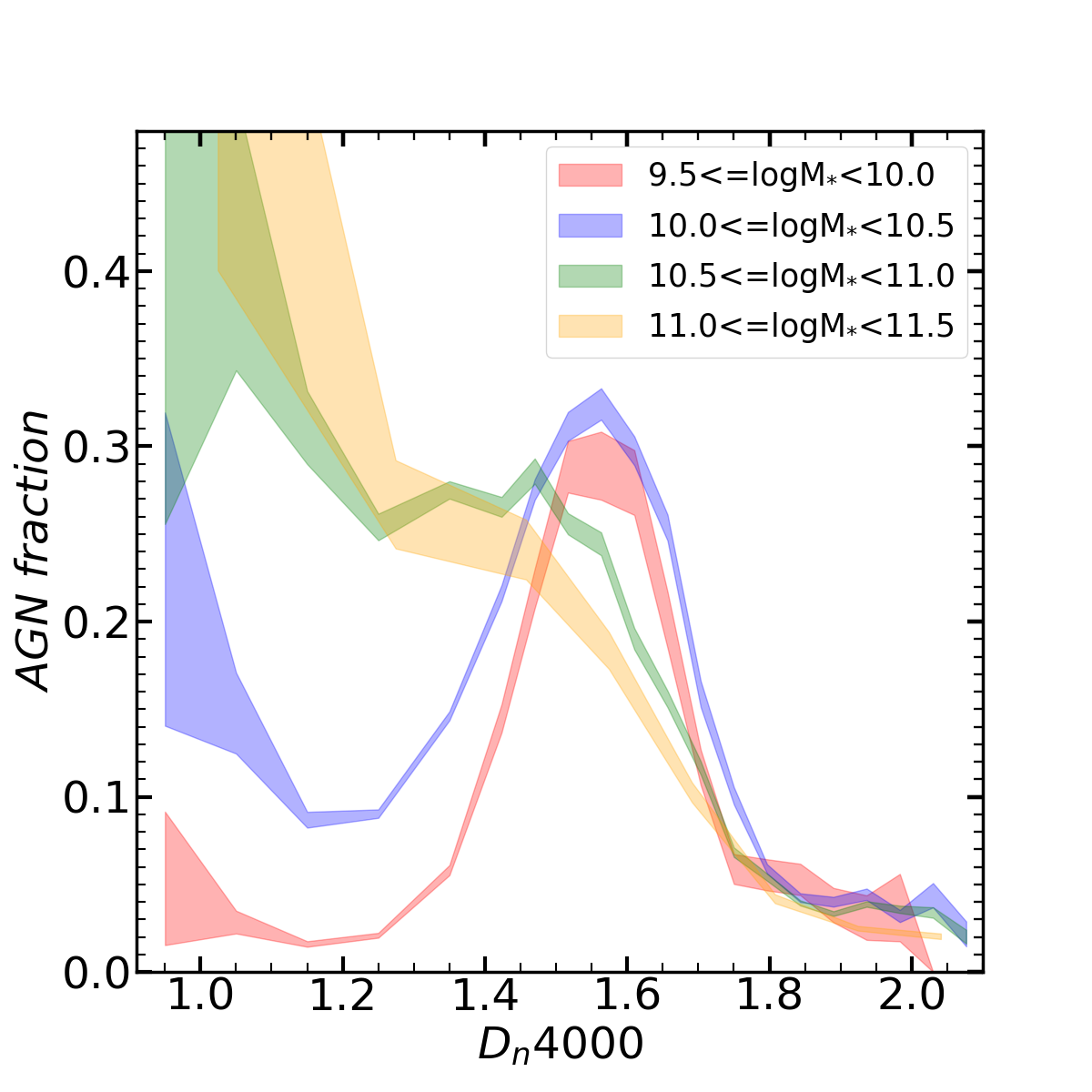}
    \caption{AGN fraction in central galaxies as a function of $D_n$4000 in four stellar mass bins as indicated by different colors. The shaded regions represent the scatter of the fraction which is calculated by using 100 bootstrap samples. Note that $V_{\rm max}$ correction is used in calculating the fraction.  See text for details.}\label{fig_AGNfrc}
\end{figure}

Figure \ref{fig_galpro} shows the probability distribution functions 
(PDFs) of the specific star formation rate (sSFR), color 
(as indicated by $(g-r)^{0.1}$), $D_n4000$
and $\sigma_{*}$ separately for AGN host galaxies, 
star-forming and quiescent galaxies, in four stellar mass bins. 
Here results are shown for cAGN (central AGNs) and the two 
control samples, cSF$^{\rm c}$ and cQ$^{\rm c}$, as defined in 
Section \ref{sec_sample}. By definition, quiescent galaxies have 
lower sSFR than star-forming galaxies, and the two populations 
have almost no overlap in their sSFR distributions within individual 
stellar mass bins. Because of the strong correlation of sSFR with 
color and $D_n4000$, quiescent galaxies have higher $(g-r)^{0.1}$ 
and $D_n4000$ than star-forming galaxies. Quiescent galaxies 
also have larger $\sigma_{*}$ than star-forming galaxies of the 
same stellar mass, consistent with the fact that the fraction of
quiescent galaxies increases rapidly with $\sigma_{*}$ 
\citep{Bluck-16, WangE-18a}. In addition, the color and $D_n4000$
for both populations increase gradually with stellar mass, 
because galaxies of lower masses in general are younger and metal
poorer.

\begin{figure}
    \centering
    \includegraphics[scale=0.22]{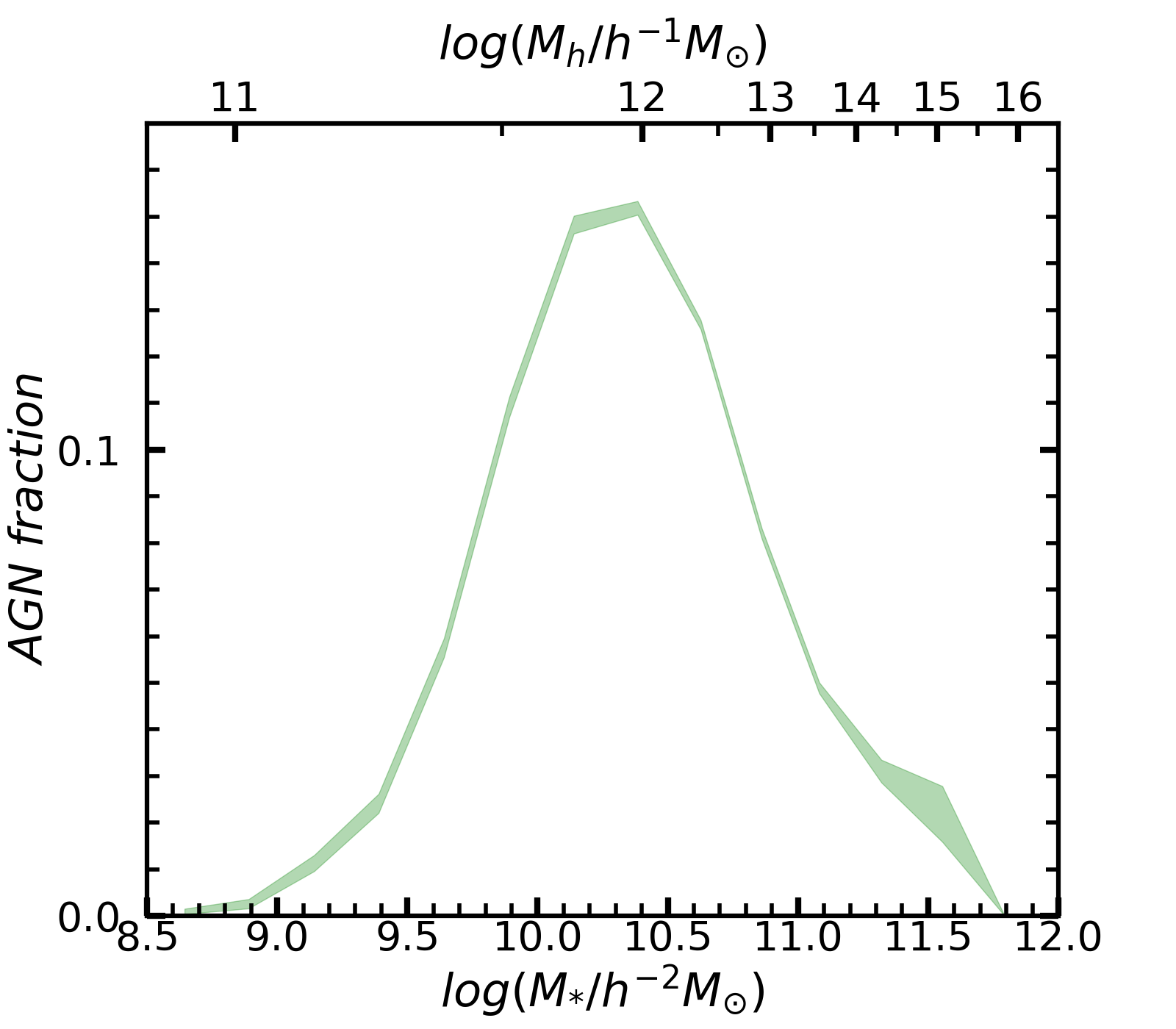}
    \caption{$V_{\rm max}$ corrected AGN fraction in central galaxies as a function of stellar mass. The top axis is the corresponding halo mass inferred by the stellar mass - halo mass relation in \cite{Yang-Mo-vandenBosch-09}. The shaded region represents the bootstrap error estimated by 100 samples.}
    \label{fig_AGN_frac_sm}
\end{figure}

  Compared to the control star-forming and quiescent galaxies, 
AGN host galaxies have a broad sSFR distribution that 
extends to both star-forming and quiescent regions. However, 
the interpretation of this result is not straightforward, 
because the SFR estimates for AGNs may have larger uncertainties
\citep[see][for details.]{Brinchmann-04}.
The velocity dispersion distribution for AGNs is between the 
two control samples, suggesting that the supermassive 
black hole (SMBH) mass ($M_{\rm BH}$) and the bulge mass of 
AGN hosts lie between the star-forming and quiescent populations. 
The difference between the AGN host galaxies and the star-forming 
galaxies becomes smaller as stellar mass increases. The color and $D_n4000$ 
distributions show similar trends, with the AGN host galaxies 
lying between the star-forming and quiescent populations. 
These results are in broad agreement with previous 
investigations \citep[e.g.][]{Man19, Dodd2020}, which 
found that AGN host galaxies tend to be in the green valley. 

 Inspecting the PDFs in details, one can notice some 
interesting features. As a reference, the vertical 
dashed lines indicate $(g-r)^{0.1}=0.8$ and $D_n4000=1.5$ 
in different panels of stellar mass bins. One can see that the
peak positions of the $(g-r)^{0.1}$ and $D_n4000$ distributions 
for AGN host galaxies are almost independent of stellar mass
over the range $9.5< \log(M_*/M_\odot)<11$. The only exception
is for the most massive galaxies, where AGN hosts on average 
are redder and have larger $D_n4000$ than their lower-mass 
counterparts. 
The difference is likely produced by the rise of a sub-population 
that has stellar populations similar to quiescent galaxies. 
This sub-population can also be seen in the other three mass bins, 
albeit less prominent.  Thus, there seem to be two 
different AGN populations, at least for massive galaxies.   
One has color and $D_n4000$ similar to quiescent galaxies, 
and this population becomes important for AGNs hosted 
by massive galaxies. The other population, which is a dominant AGN population, 
has color and $D_n4000$ distributions  
that are independent of stellar mass. Note that for normal galaxies
both the color and $D_n4000$ distributions shift 
to the redder and higher-$D_n40000$ sides with increasing stellar mass, 
and the trend is particularly strong for star-forming galaxies. 
The mass-independence of the color and $D_n4000$ distributions  
for the AGN population with $\log(M_*/M_\odot)<11$ thus 
indicates that AGN host galaxies do not always lie in between star-forming and quiescent galaxies.
It is likely that AGN hosts experienced a specific stage.

Figure \ref{fig_AGNfrc} shows the AGN fraction as a function of $D_n4000$
in four stellar mass bins. Here, the AGN fraction is calculated by using the whole
central galaxy sample (cG) with $V_{\rm max}$ weighting
\citep{Blanton-Roweis-07} and correction for redshift incompleteness 
\citep{Blanton-05a}. The results clearly show two peaks, one 
at $D_n4000\sim1.0$ and the other at $D_n4000\sim1.5$. 
The peak height at $D_n4000\sim 1$ depends strongly on stellar mass:
for galaxies with $\log(M_*/\msunt)>10$, the AGN fraction 
is about 20\% to 40\%, and the fraction declines to 5\% at the lowest 
stellar mass bin. In contrast, the second peak 
depends only weakly on stellar mass, with a value of about 30\%.
We show the AGN fraction as a function of stellar mass for central
galaxies in Figure \ref{fig_AGN_frac_sm}. AGN fraction 
is a strong function of stellar mass and peaks at
stellar mass of about $10^{10.4}\msunt$. 
The lower fraction at low (high) mass end reflects 
that these galaxies are dominated by star-forming and small 
$D_n4000$ (quiescent and high $D_n4000$) galaxies. 
The mean AGN fraction in the four stellar mass bins are about 8\%, 15\%, 11\% 
and 5\%, respectively. The fractions in the two 
$D_n4000$ peaks are much higher than the mean values, 
suggesting that AGNs tend to be hosted by 
galaxies in some specific evolution stages. 

The low value, $D_n4000\sim 1$, of the first peak signifies 
the existence of a very young stellar population in the central parts of the host galaxies.  
\footnote{Note that SDSS fiber size limits the aperture 
over which the light from a galaxy is collected.} 
As shown in \citet{Kauffmann-03} and \citet{Greene2020}, 
the stellar age corresponding to $D_n4000\sim 1$ is 
typically smaller than $10^8$ years, indicating that 
the stars in the central parts of these galaxies formed 
through short bursts. Because of the short time scale, galaxies 
observed with such a young stellar population are rare, 
which may explain the absence of the peak in the PDFs shown 
in Figure \ref{fig_galpro}. The peak at $D_n4000=1$, 
therefore, suggests that galaxies with strong current star formation
have a strong tendency to be AGNs hosts. This is consistent with 
the result of \cite{Greene2020}, who found that the fraction of AGN hosts 
among star-burst galaxies is high, and suggests that the process 
associated with a star burst may trigger AGN activities.

The second peak at $D_n4000\sim 1.5$ corresponds to the
dominant AGN population shown in Figure \ref{fig_galpro}. 
This peak was not found in \cite{Greene2020}, because they only
focused on star burst galaxies and their sample contained 
only few galaxies with $D_n4000\ge1.5$.
\cite{Kauffmann-AGN-03} studied the H$\delta$ absorption 
lines of AGN host galaxies and found that a significant fraction 
of them have experienced a star burst phase within the past 
1-2 Gyrs.  Thus, the host galaxies of AGNs in this peak may 
have also been triggered by a processes associated with a star burst.
However, since the life time for AGN activities is believed to be less
than $10^8$ years\citep[e.g.][]{Marconi-04, Schawinski15, Yuan18}, the observed AGNs in this peak 
cannot be directly related to the star bursts that happened 1 to 2 
Gyrs ago. We will come back to the implications of this results later.

\begin{table}
    \caption{Halo masses derived using lensing and clustering measurements for AGNs and their control samples.}\label{tab_01}
    \centering
    \begin{tabular}{c|c|c|c}
    \hline\hline
    Sample name & $\log M_*$ &  $\log M_{\rm h}$ (M1) & $\log M_{\rm h}$(M2)  \\
    \hline
    cAGN  &    & 11.85$^{+0.15}_{-0.19}$ & 11.96$^{+0.06}_{-0.07}$ \\
    cSF$^{\rm c}$ & All  & 11.85$^{+0.07}_{-0.08}$ & 11.81$^{+0.06}_{-0.07}$\\
    cQ$^{\rm c}$ &   & 12.38$^{+0.1}_{-0.06}$ & 12.39$^{+0.07}_{-0.05}$\\
    \hline

    cAGN  &  [9.5,10.0]  &  11.68$^{+0.42}_{-0.44}$ & 11.63$^{+0.23}_{-0.26}$ \\
      &  [10.0,10.5]     & 11.45$^{+0.32}_{-0.3}$         & 11.68$^{+0.1}_{-0.11}$\\
      &  [10.5,11.0]     & 12.08$^{+0.2}_{-0.27}$         & 12.1$^{+0.08}_{-0.08}$\\
      &  [11.0,11.5]     & 12.85$^{+0.23}_{-0.3}$         & 12.91$^{+0.12}_{-0.12}$\\
    \hline
    cSF$^{\rm c}$  &  [9.5,10.0]  &  11.85$^{+0.22}_{-0.34}$  & 11.39$^{+0.3}_{-0.26}$ \\
      &  [10.0,10.5]    &  11.69$^{+0.15}_{-0.2}$         & 11.64$^{+0.11}_{-0.13}$\\
      &  [10.5,11.0]    &  11.9$^{+0.13}_{-0.16}$         & 11.88$^{+0.1}_{-0.11}$\\
      &  [11.0,11.5]    &  13.12$^{+0.12}_{-0.14}$        & 12.99$^{+0.1}_{-0.11}$\\
    \hline
    cQ$^{\rm c}$  &  [9.5,10.0]   &  11.83$^{+0.33}_{-0.45}$  & 12.17$^{+0.16}_{-0.18}$ \\
      &  [10.0,10.5]    &  12.14$^{+0.11}_{-0.1}$         & 12.16$^{+0.07}_{-0.07}$\\
      &  [10.5,11.0]    &  12.43$^{+0.07}_{-0.08}$        & 12.44$^{+0.06}_{-0.06}$\\
      &  [11.0,11.5]    &  13.11$^{+0.11}_{-0.1}$         & 13.22$^{+0.12}_{-0.09}$\\
    \hline\hline
    cAGN$_{\sigma_{*}}$     &   & 11.75$^{+0.18}_{-0.23}$  & 11.92$^{+0.08}_{-0.08}$ \\
    cSF$^{\rm c2}_{\sigma_{*}}$  &  All  & 11.73$^{+0.09}_{-0.09}$  & 11.76$^{+0.08}_{-0.08}$ \\
    cQ$^{\rm c2}_{\sigma_{*}}$   &    & 12.22$^{+0.13}_{-0.08}$  & 12.32$^{+0.08}_{-0.07}$\\
    \hline
    cAGN${\sigma_{*}}$  &  [9.5,10.0]  &  11.63$^{+0.43}_{-0.41}$ & 11.69$^{+0.22}_{-0.25}$ \\
                             &  [10.0,10.5] &  11.44$^{+0.31}_{-0.29}$ & 11.73$^{+0.11}_{-0.12}$\\
                             &  [10.5,11.0] &  12.04$^{+0.21}_{-0.31}$ & 12.13$^{+0.08}_{-0.08}$\\
                             &  [11.0,11.5] &  12.77$^{+0.26}_{-0.4}$  & 13.12$^{+0.11}_{-0.11}$\\
    \hline
    cSF$^{\rm c2}_{\sigma_{*}}$  &  [9.5,10.0]  &  11.69$^{+0.34}_{-0.41}$ & 11.47$^{+0.3}_{-0.29}$ \\
                            &  [10.0,10.5] &  11.72$^{+0.13}_{-0.16}$ & 11.72$^{+0.11}_{-0.12}$\\
                            &  [10.5,11.0] &  11.97$^{+0.12}_{-0.14}$ & 11.97$^{+0.09}_{-0.1}$\\
                            &  [11.0,11.5] &  13.23$^{+0.08}_{-0.1}$  & 13.4 $^{+0.09}_{-0.1}$\\
    \hline
    cQ$^{\rm c2}_{\sigma_{*}}$  &  [9.5,10.0]  &   11.97$^{+0.28}_{-0.44}$ & 12.21$^{+0.15}_{-0.18}$ \\
                           &  [10.0,10.5] &   12.11$^{+0.14}_{-0.12}$ & 12.16$^{+0.09}_{-0.09}$\\
                           &  [10.5,11.0] &   12.27$^{+0.09}_{-0.1}$ &  12.32$^{+0.07}_{-0.07}$\\
                           &  [11.0,11.5] &   13.07$^{+0.12}_{-0.14}$ & 13.15$^{+0.09}_{-0.1}$\\
    \hline\hline
    \end{tabular}
\end{table}

\section{Masses of AGN Host Halos}\label{sec_hmng}

In Section \ref{sec_ssm}, we described how one can estimate 
the host halo mass of central galaxies using gravitational lensing 
signals and the 2PCCF. Here we apply the methods to AGNs and the corresponding 
control samples of normal galaxies to investigate the masses of AGN  host halos 
in comparison to those of normal galaxies.

\subsection{Results from galaxy-galaxy lensing}
\label{sec_mhgglens}

\begin{figure*}
    \centering
    \includegraphics[scale=0.72]{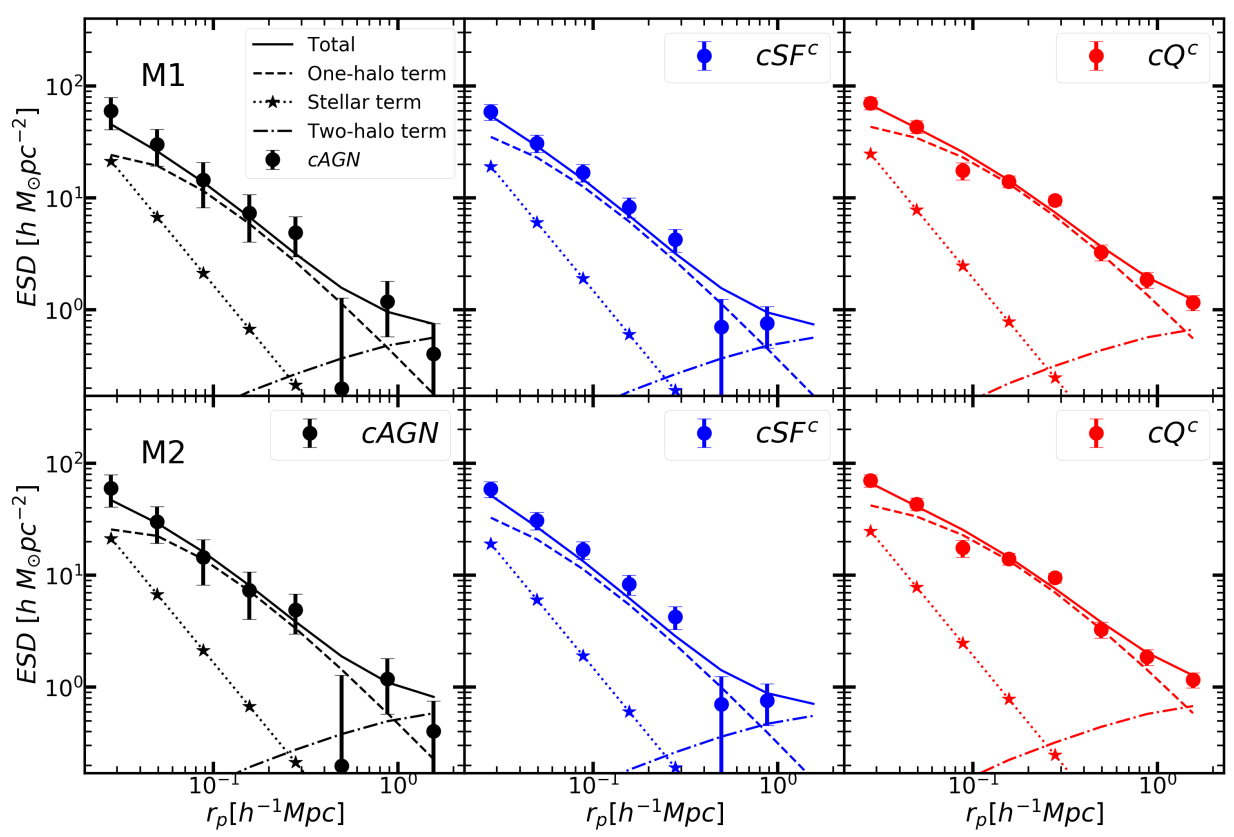}
    \caption{The lensing signal and the best fitting results for cAGN (black, left column), cSF$^{\rm c}$ (blue, middle column) and cQ$^{\rm c}$ (red, right column) with different methods (the upper row is for M1, the lower row for M2). In each panel, the dots with error bars are the lensing signal, while the dashed line, dotted line with stars and point line represent contribution from one-halo term, stellar mass term and two-halo term, respectively. The total fitting result is indicated by the solid line. }
    \label{fig_lensing}
\end{figure*}

\begin{figure*}
    \centering
    \includegraphics[scale=0.71]{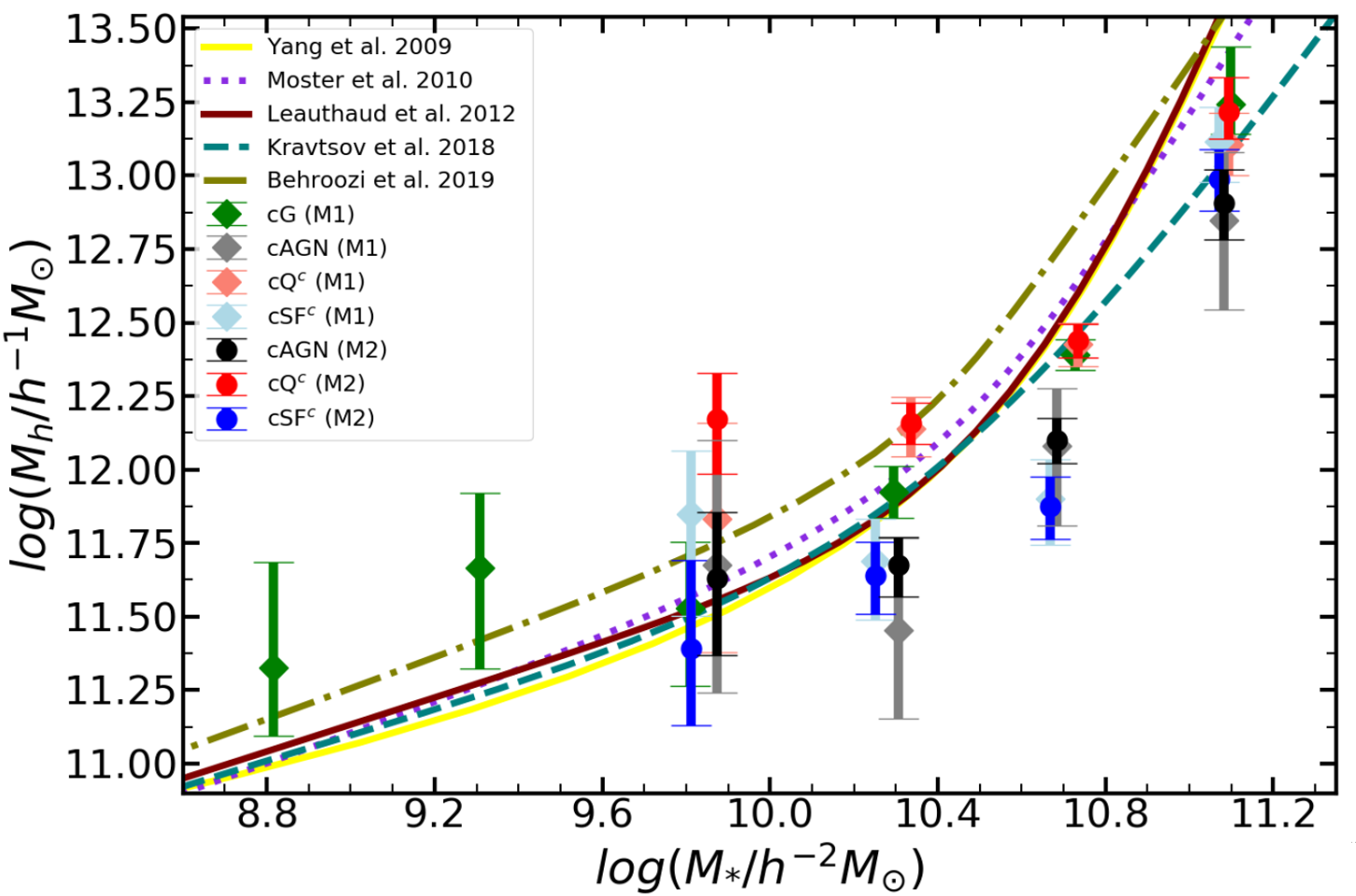}
    \caption{Stellar mass-halo mass relation for central AGNs and their control galaxies. For our results, those halo masses derived by M1 are given by diamonds with error bars, while those derived by M2 are given by dots with error bars. For comparison, we also show the SHMR in the literature, obtained by using various methods, including galaxy group catalog\citep{Yang-Mo-vandenBosch-09}, abundance matching\citep[see e.g.][]{Moster10, Behroozi-19}, conditional luminosity function\citep{Kravtsov2018}, and weak lensing \citep{Leauthaud-12}. }
    \label{fig_shmr}
\end{figure*}

In Figure \ref{fig_lensing}, we show the excess surface density
profiles derived from the g-g lensing signal for sample cAGN 
and corresponding control samples cSF$^{\rm c}$ and cQ$^{\rm c}$. As one can see, the ESD 
profiles obtained from cAGN and cSF$^{\rm c}$ are quite similar, while 
that for cQ$^{\rm c}$ is higher. To quantify the results, 
We use model M1 (see Section \ref{sec_gg} and also \cite{Luo-18}) 
to fit the observed ESD profiles and to derive an average halo mass for each of the three 
samples. The results of the halo mass are listed in Table \ref{tab_01}. 
The halo mass for central AGNs is about $10^{11.85}\msun$, 
in agreement with the g-g lensing results for AGNs selected from the SDSS DR4 
\citep[e.g.][]{Mandelbaum09}. The halo mass for the control sample of 
star-forming galaxies, cSF$^{\rm c}$, is very similar to that of AGNs, while the mean 
halo mass for the quiescent galaxies, about $10^{12.38}\msun$, is about 
three times as high as those for AGNs and star-forming galaxies
of the same stellar mass. We have also carried out the same analysis
for AGN hosts and normal galaxies in four stellar mass bins, and corresponding results 
are listed in Table \ref{tab_01} and plotted in Figure.\ref{fig_shmr}. 
As expected, for each population, the average halo mass is larger for galaxies 
with larger stellar masses. For a given stellar mass, the average halo
masses for AGN hosts and star-forming galaxies are similar but lower than 
that of quiescent galaxies. 

Because of the limit by the sample size, the halo masses obtained from the 
g-g lensing measurements are quite uncertain, particularly when galaxies 
are divided into sub-samples of stellar mass. However, additional constraints
on halo mass can also be obtained through halo bias estimated from 
the clustering strength on large scales. In what follows, we present results 
based on the clustering measurements.

\subsection{Constraints from 2PCCF}
\label{sec_mh2PCCF}

\begin{figure*}
    \centering
    \includegraphics[scale=0.65]{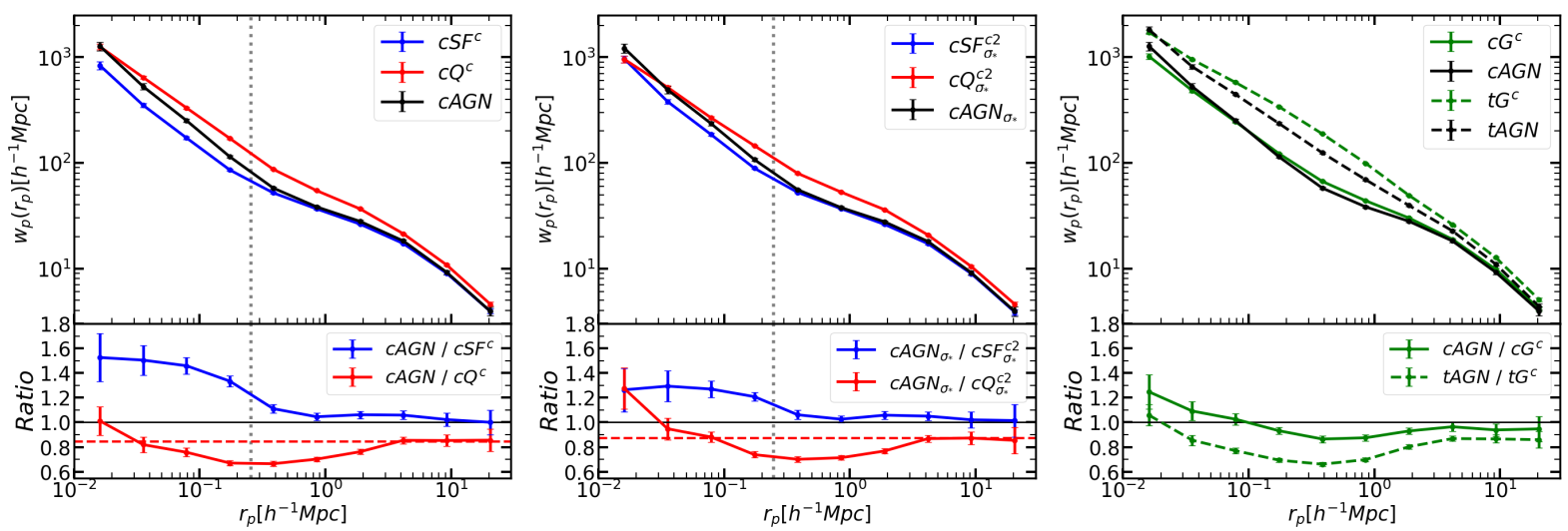}
    \caption{Left panels: the upper panel shows the 2PCCF of cAGN (black) and its control star-forming (cSF$^{\rm c}$) and quiescent galaxies (cQ$^{\rm c}$). The lower panel shows the 2PCCF ratio of cAGN to cSF$^{\rm c}$ (blue) and cAGN to cQ$^{\rm c}$ (red). The horizontal dashed line (red) indicates the theoretical halo bias \citep{Tinker-10} ratios between cAGN and cQ$^{\rm c}$ with halo mass derived by M1. The vertical dotted line (grey) indicates the virial radius of the host halo of cAGN derived by M2. Middle panels: Similar to the left panels, but 
    for cAGN$_{\sigma_*}$ and their control galaxies, cSF$^{\rm c2}_{\sigma_*}$ and cQ$^{\rm c2}_{\sigma_*}$. Right panels: upper panel shows the 2PCCF for cAGN and its control galaxies (cG$^{\rm c}$), tAGN and its control galaxies (tG$^{\rm c}$). The lower panel shows the 2PCCF ratio of cAGN to cG$^{\rm c}$ and tAGN to tG$^{\rm c}$. Please see Section \ref{sec_sample} for the sample construction.}\label{fig_2pcf}
\end{figure*}

\begin{figure*}
    \centering
    \includegraphics[scale=0.85]{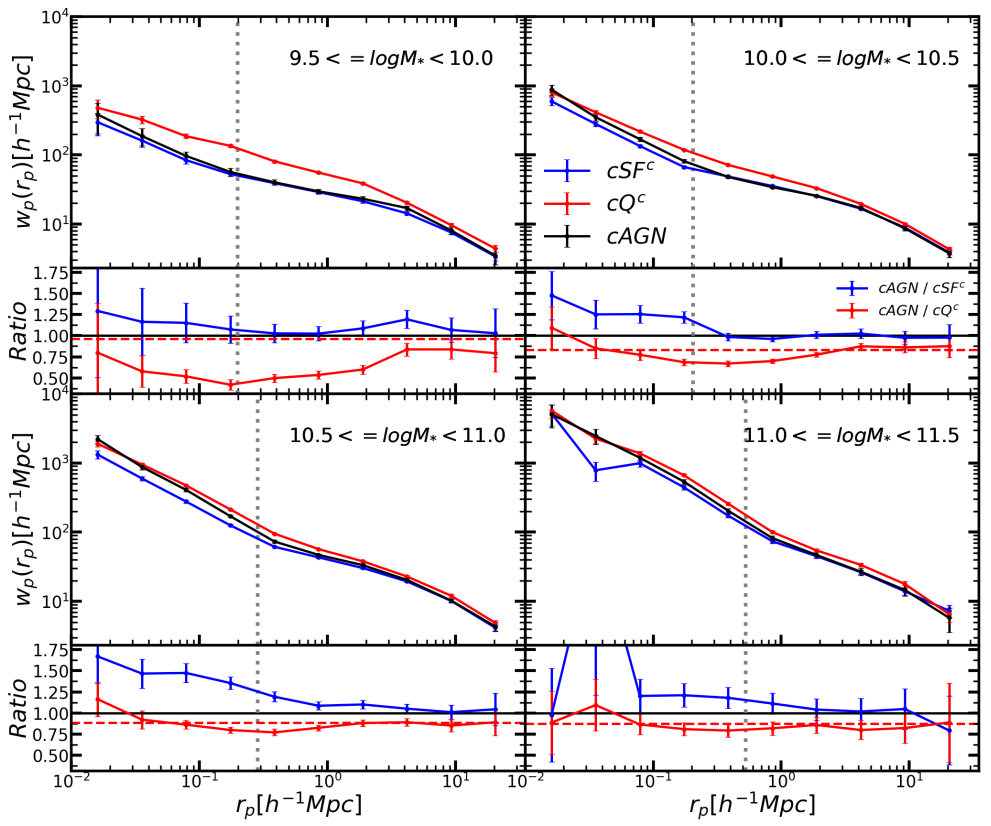}
    \caption{Similar to the left panels in Figure \ref{fig_2pcf}, but cAGN, cQ$^{\rm c}$ and cSF$^{\rm c}$ are split into four stellar mass bins as indicated in each panel.}
    \label{fig_st}
\end{figure*}

Figure \ref{fig_2pcf} shows the projected 2PCCFs,
$w_{\rm p}(r_{\rm p})$, for central AGNs (cAGN) in comparison to 
the corresponding control galaxies.
The 2PCCFs of the three samples exhibit some interesting features on both 
small and large scales. We will talk about the small-scale in 
the next section; here we focus on the properties on large scales in 
connection their implications for halo masses. As one can see, AGNs
have almost the same clustering amplitude as the control star-forming
galaxies at scales larger than about $0.4\mpc$, suggesting that the 
host halos for the two populations have very similar large-scale bias
and halo mass. At scales larger than $\sim 4\mpc$, the cAGN/cQ$^{\rm c}$ 
ratio is almost a constant and is less than one, indicating that 
quiescent galaxies reside in more massive halos than both 
AGNs and star-forming galaxies. We have also estimated the 2PCCF results 
for galaxies in the same four stellar mass bins as used in the g-g 
lensing analysis, and the results are shown in Figure \ref{fig_st}. 
The results are consistent with those shown in Figure \ref{fig_2pcf}. 
At large scales, cQ$^{\rm c}$ are more strongly clustered than both 
cAGN and cSF$^{\rm c}$, and cAGN has the same clustering strength as cSF$^{\rm c}$. 

The results obtained from the 2PCCFs on large scales are 
thus consistent with the interpretation of the g-g lensing 
results in terms of halo mass. Using the halo masses derived 
from the lensing (M1 method) and the theoretical model for 
halo bias described in \cite{Tinker-10}, we can predict 
the ratio of the 2PCCF on large scales between cAGN and cQ$^{\rm c}$. 
The ratios for different cases are shown as the horizontal 
dashed lines in Figure \ref{fig_2pcf} and the corresponding panels 
of Figure \ref{fig_st}. We see that the lensing 
and clustering results are in good agreement 
for the three high mass bins. The discrepancy for the 
lowest mass bin is difficult to judge, as the uncertainties 
for both measurements are large.

The good agreement between the lensing and clustering results suggests
that we can combine the results to obtain tighter constraints 
on halo masses using model M2 described in Section \ref{sec_gg}.
Since cAGN and cSF$^{\rm c}$ have a similar cross correlation amplitude 
on scales larger than $\sim 1\mpc$, and the ratio of 
the 2PCCF between cAGN and cQ$^{\rm c}$ is roughly a constant at
scales larger than $\sim 4\mpc$, the likelihood terms for the 2PCCF 
(Equation \ref{eq_ratio}) are calculated using the ratios at 
$r_{\rm p}>1\mpc$ for cAGN$/$cSF$^{\rm c}$ and at $r_{\rm p}>4\mpc$ for 
cAGN$/$cQ$^{\rm c}$. For comparison, the best-fitting models to the ESD 
profiles are shown in the lower panels of Figure \ref{fig_lensing}, and 
the derived halo masses are given in Table \ref{tab_01}.
As one can see, the halo masses derived from model M2 agree very well 
with those from M1, indicating again that the lensing and clustering 
results are consistent with each other. The combined constraints
also lead to smaller uncertainties, as expected. 

The stellar mass - halo mass relation (SHMR) obtained from model M2 
is shown in Figure \ref{fig_shmr}. For comparison, the result 
for the total central sample, cG, obtained using model M1 is shown 
as green points.  As references, the SHMR derived by 
various methods in the literature, including galaxy group 
catalog \citep{Yang-Mo-vandenBosch-09}, abundance matching 
\citep[see e.g.][]{Moster10, Behroozi-19}, 
conditional luminosity function \citep{Kravtsov2018} and
weak lensing \citep{Leauthaud-12} are presented. Our result 
for the total sample is in good agreement with previous results, 
indicating that our method is reliable. 
In general, the halo mass increases with stellar mass.
And there is a pivot halo mass around $10^{12}\msun$, 
above and below which the SHMR have different slopes. 

Our analysis, combining weak lensing and clustering measurement, 
clearly show that, at given $M_*$, the host halos of 
quiescent galaxies are more massive than those of 
star-forming galaxies and AGN host galaxies. 
The difference is particularly significant in the two middle 
$M_*$ bins, which include most (about 88\%) of the AGNs.
It is in agreement with previous studies that found
quiescent galaxies reside in more massive halos than star-forming galaxies 
of the same $M_*$\citep[e.g.][]{Mandelbaum06, Behroozi-19}.
It is broadly consistent with the passive quenching model\citep{Wechsler-Tinker18}, 
in which star-forming galaxies grow faster than quiescent galaxies, 
while their host halos grow in a statistically similar manner.
The host halo masses of AGN host are in good agreement 
with those for star-forming galaxies, indicating that the 
two populations of galaxies may be connected, as we will 
discuss in Section \ref{sec_dis}.

\section{Satellite Galaxies around AGNs}\label{sec_sat}

In the last section, we have shown that the masses of 
AGN halos are similar to those of star-forming galaxies of the 
same stellar mass. In this section, we examine further whether  
or not the host halos of AGNs and star-forming galaxies may be different 
in the number of satellite galaxies they contain. The answer to 
this question may shed light on the roles of galaxy-galaxy interaction 
in triggering AGN activities. In the literature, there are suggestions 
that the AGN activities may depend on the central velocity dispersion 
of galaxies. We therefore also check whether or not the satellite abundance 
depends on the central velocity dispersion of the central galaxies.  

\subsection{Excess of satellites around AGNs}
\label{sec_exst}

\begin{figure*}
    \centering
    \includegraphics[scale=0.51]{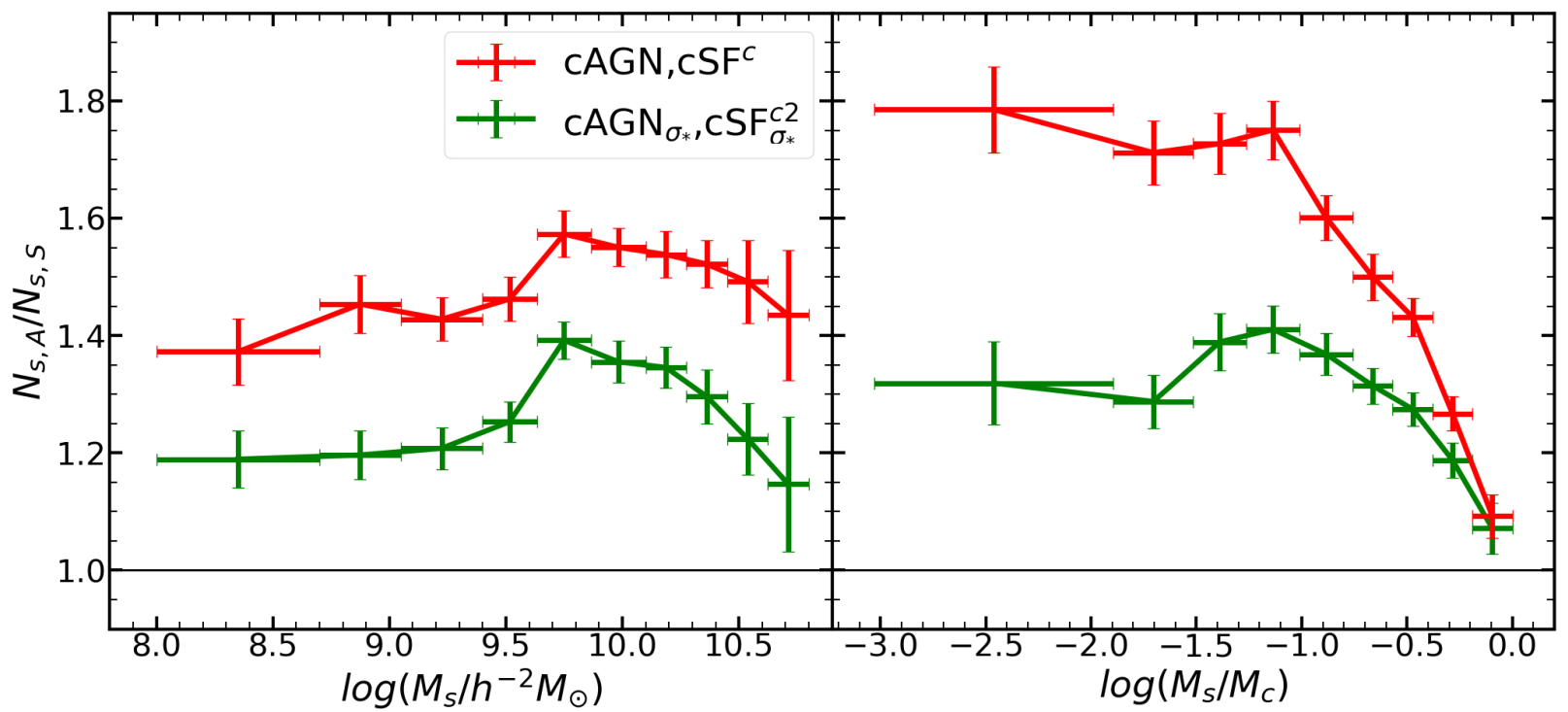}
    \caption{The left panel shows the ratio of the mean satellite number around two cAGN samples (cAGN and cAGN$_{\sigma_{*}}$ ) over the mean satellite number around two corresponding control star-forming samples (cSF$^{\rm c}$ and cSF$^{\rm c2}_{\sigma_{*}}$) as function of satellite stellar mass, respectively. The right panel is similar to the left, but the ratio is as function of satellite stellar mass ($M_{\rm s}$) over central stellar mass ($M_{\rm c}$).}\label{fig_msmc}
\end{figure*}

As shown in the left panels of Figure \ref{fig_2pcf}, 
AGNs are more strongly clustered than star-forming galaxies at small
scales, although both populations have similar 2PCCF on large scales. 
At $r_{\rm p}< 300\kpc$, the ratio in 2PCCF between cAGN and cSF$^{\rm c}$ 
increases and reaches about 1.5. The mean halo mass for 
the two samples, which is about $10^{11.96}\msun$, corresponds 
to a mean halo virial radius, $r_{\rm vir}\approx 0.26\mpc$,
and is indicated by the vertical dotted line. As one can see, 
the virial radius separates the 2PCCF into two distinct parts.  
The slope of $w_{\rm p}(r_{\rm p})$ becomes 
much steeper at scales smaller than the virial radius, and 
this is true for both the AGN and star-forming samples.
This reflects the transition of the correlation function 
from one-halo to two-halo terms, providing an additional
support to the reliability of our halo mass estimate.  
Within the virial radius, the cross correlation strength 
for AGNs is enhanced relative to that for star-forming galaxies, 
indicating that the average number of satellites around AGNs 
is higher than that around central star-forming galaxies.
+
 The results shown in Figure \ref{fig_st} for galaxies in 
four different stellar mass bins indicate that the enhancement 
of the correlation strength within the virial radius for AGNs   
is present in the two intermediate mass bins. In the lowest mass bin, 
there are only 3,604 cAGNs, making the error bars quite large. 
In the most massive bin, the enhancement seems to be insignificant.
This might owe to the fact that a large fraction of the AGN host galaxies 
in this mass bin are quiescent galaxies, different from the 
main AGN population (see Section \ref{sec_galpro}).
However, because of the limited sample size (the number of cAGNs in the 
most massive bin is 1,634), we are not able to obtain a definite 
answer to this question. Based on the halo mass estimated 
for individual subsamples, we derive the corresponding virial 
radii and show them as the vertical dotted lines. 
Similar to Figure \ref{fig_2pcf}, the change of the slope of 
the 2PCCFs occurs around the virial radii and the enhancement 
of the 2PCCF for AGNs is found only within the virial radii. 

To examine which kind of satellites contribute to the difference
between AGNs and star-forming galaxies, we identify satellites 
around galaxies in cAGN and cSF$^{\rm c}$ as follows. 
For a given central AGN in cAGN or a star-forming galaxy
in cSF$^{\rm c}$, we select satellites from the reference galaxy 
sample according to the following criteria, $|\delta z|\leq 3 v_{\rm vir}/c$,
$r_{\rm p}\leq r_{\rm vir}$ and $M_{\rm s}< M_{\rm c}$.
Here $r_{\rm vir}$ and $v_{\rm vir}$ are, respectively, the virial radius
and virial velocity calculated assuming a halo mass of $10^{11.96}\msun$,
$\delta z$ is the redshift difference between a central and a satellite,
and $M_{\rm c}$ and $M_{\rm s}$ are the stellar masses
of the central and satellite, respectively. We then compute the mean 
numbers of satellites around the central AGNs and 
star-forming galaxies, denoted by $N_{\rm s,A}$ and $N_{\rm s, S}$,  
respectively. The errors are calculated by using 100 bootstrap samples. Due to Malmquist bias, $N_{\rm s,A}$ or $N_{\rm s, S}$ 
does not represent the true number of satellites around the centrals. 
However, since cAGN and cSF are matched in redshift, the Malmquist 
bias is expected to have a similar effect on both so that the   
number ratio, $N_{\rm s,A}/N_{\rm s,S}$, should not be affected 
significantly. Figure \ref{fig_msmc} shows $N_{\rm s,A}/N_{\rm s,S}$
as a function of $M_{\rm s}$ (left panel) and $M_{\rm s}/M_{\rm c}$
(right panel). The ratio, $N_{\rm s,A}/N_{\rm s,S}$, 
changes little with the satellite mass,  
except the small peak around $\log(M_{\rm s}/\msunt)=9.7$. 
In contrast, the ratio changes strongly with $M_{\rm s}/M_{\rm c}$.
The mean number of satellites around AGNs is similar 
to that around star-forming galaxies for satellites with 
stellar masses comparable to the centrals ($M_{\rm s}/M_{\rm c}\sim 1$).
As $M_{\rm s}/M_{\rm c}$ decreases, the number ratio increases
rapidly to $\sim 1.8$ at $M_{\rm s}/M_{\rm c}=0.1$ 
and remains almost constant down to $M_{\rm s}/M_{\rm c}=0.001$. 
Thus, the difference in satellite abundance between central AGNs 
and star-forming galaxies is larger for satellites of lower 
masses relative to their centrals. 

To summarize, our results clearly show that central AGNs are
surrounded by more satellites, especially small
satellites with masses less than about one-tenth of the central 
mass, than star-forming galaxies in the control sample. 
This suggests that local environments, as represented 
by the abundance of satellites, play an important role
in triggering AGN activity, as we will discuss later.

\subsection{The effects of galaxy central velocity dispersion}\label{sec_vd}

In above sections, control samples are matched with AGN samples
only in stellar mass and redshift. Since central velocity dispersion 
of galaxies, which is related to the mass of the central bulge, 
has been suggested as an important parameter related to galaxy 
quenching, it is interesting to investigate its impacts on our results.  
To this end, we examine cAGN$_{\sigma_*}$ in comparison with the 
corresponding control samples, cSF$^{\rm c2}_{\sigma_*}$ and cQ$^{\rm c2}_{\sigma_*}$.
As detailed in Section \ref{sec_sample}, these samples are constructed 
by matching them not only in stellar mass and redshift, but also 
in central velocity dispersion, $\sigma_*$. 
The analyses for these samples 
are the same as those described above.
The average halo masses estimated from g-g lensing 
measurements using Models M1 and M2 are listed in Table \ref{tab_01}, and 
the results about the projected 2PCCFs are shown in the middle panel of Figure \ref{fig_2pcf}.

As one can see, AGNs and control star-forming galaxies still have 
similar mean halo mass and clustering amplitude at large scales, 
even after $\sigma_*$ is constrained. Here again, quiescent galaxies 
tend to reside in more massive halos and are more strongly clustered. 
The halo mass obtained from lensing measurements is consistent 
with that based on the clustering results, as shown by the horizontal dashed 
line, which indicates the prediction of the halo bias model using the 
lensing mass. In fact, controlling $\sigma_*$ does not change the 
results on halo mass significantly (Table \ref{tab_01}).

On small scales, $r_{\rm p}< r_{\rm vir}$, 
we can see a clear change in the slope of 
$w_{\rm p}(r_{\rm p})$ at $r_{\rm p}\sim r_{\rm vir}$,
and a significant excess in clustering strength for AGNs relative to
the control star-forming galaxies. Thus, consistent with the 
results presented earlier, AGNs are surrounded by more satellites 
than are star-forming galaxies. However, the excess becomes smaller, 
with the ratio reduced to about 1.3, after $\sigma_*$ is controlled.
Figure \ref{fig_msmc} shows $N_{\rm s,A}/N_{\rm s,S}$
as a function of $M_{\rm s}$ and $M_{\rm s}/M_{\rm c}$
for the new samples. Although the overall trend
is similar to that shown above, the ratio is smaller. 
For example, at $\log M_{\rm s}/M_{\rm c}= -1.0$,
the mean number of satellites around AGNs is about 1.4
times that around star-forming galaxies.

\begin{figure}
    \centering
    \includegraphics[scale=0.35]{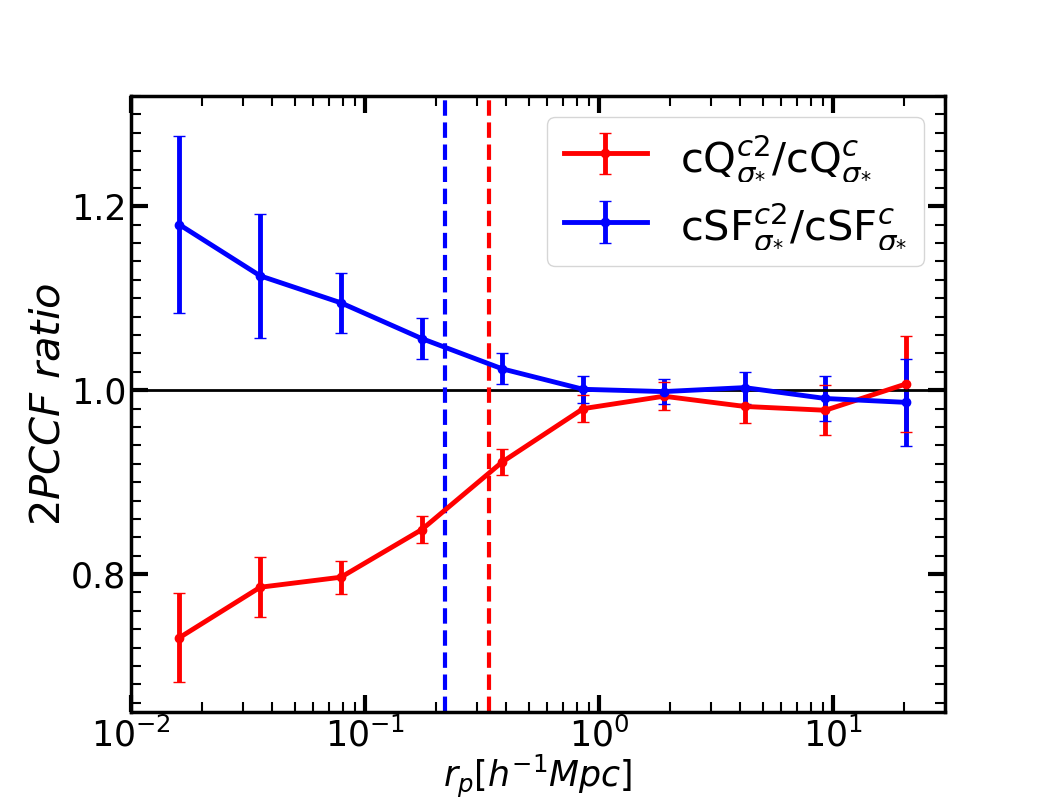}
    \caption{The 2PCCF ratio of cQ$^{\rm c2}_{\sigma_*}$/cQ$^{\rm c}_{\sigma_*}$ (red) and cSF$^{\rm c2}_{\sigma_*}$/cSF$^{\rm c}_{\sigma_*}$ (blue). The red and blue vertical dashed line represent the varial radius of host halos for cQ$^{\rm c2}_{\sigma_*}$ and cSF$^{\rm c2}_{\sigma_*}$ derived by M2, respectively. See Section \ref{sec_vd} for the details of the construction of the control samples.}
    \label{fig_2pcfvdr}
\end{figure}

To understand the new results, it is interesting to check 
whether or not galaxy clustering depends on $\sigma_*$ for fixed 
stellar mass. As shown in Section \ref{sec_sample}, there is a small 
difference between cAGN$_{\sigma_*}$ and cAGN samples.
To have a fair comparison, we construct two additional control 
samples, cSF$^{\rm c}_{\sigma_*}$ and cQ$^{\rm c}_{\sigma_*}$, 
for cAGN$_{\sigma_*}$, by only matching $M_*$ and $z$.
Figure \ref{fig_2pcfvdr} shows the 2PCCF ratios 
between the control samples with and without controlling $\sigma_*$, 
respectively. For star-forming and quiescent galaxies, controlling $\sigma_*$ 
does not change the clustering at the scales larger than the virial radius, 
consistent with our halo mass measurement. 
At scales less than the virial radius (indicated by the vertical dashed lines), 
however, cQ$^{\rm c}_{\sigma_*}$ more strongly clustered 
than cQ$^{\rm c2}_{\sigma_*}$, while cSF$^{\rm c}_{\sigma_*}$
is less strongly correlated then cSF$^{\rm c2}_{\sigma_*}$.  
Since on average the AGN sample has smaller $\sigma_*$ than the 
quiescent sample, but larger $\sigma_*$ than the  
star-forming one, as shown in Figure \ref{fig_galpro}, these two 
results indicate that, at given stellar mass, galaxies with higher 
${\sigma_*}$ are surrounded by more satellites.
This suggests that the present of satellite galaxies may play a role 
in the buildup of bulges.
It suggests that the bulge formation and AGN activities may be
caused by the same mechanism. We will come back to this issue in Section \ref{sec_fq}.

\section{Discussions and Implications}\label{sec_dis}

\subsection{Impact of the AGN sample selection}

AGNs used in our analysis are identified from the SDSS galaxy 
main sample based on narrow emission lines. 
AGNs with weak or insignificant narrow lines, which may be 
produced by low black-hole activities, are not 
included. Moreover, the majority of 
the AGNs used here are type II AGNs, in which broad lines 
and continuum are expected to be blocked by dusty structures.
Thus, type I AGNs that show broad lines are not included in our 
analysis. Furthermore, we consider only central AGNs.
So selected, our AGN sample is a biased sample of the total 
AGN population. Detailed analyses are required to examine 
the potential impact of these selections;  here we present 
some discussion about the issue.

According to the standard unified model \citep{Antonucci93}, 
the difference between type Is and IIs are 
attributed solely to orientation effects. 
In this case, ignoring type Is should not significantly 
change our results. However, there is growing evidence against 
the unified model. For example,  
\citet{Jiang2016} and \citet{Powell2018} 
found that type Is have the same clustering as type IIs at large scales,
but have significantly weaker clustering at small scales,
indicating that they reside in halos of similar mass but 
with different satellite abundances. Thus, if type I AGNs were 
included in the analysis, the difference in small-scale 
clustering between AGNs and star-forming galaxies presented
above would be reduced. 

The typical $\rm [OIII]$ luminosity of our type II AGN sample 
is less than $10^{41} \rm erg~s^{-1}$, at which the 
type I fraction is less than 20\%\citep{Simpson2005, Khim2017}. 
This means that Type Is can only be a small part of the whole AGN sample, 
and that the impact of excluding type Is may not be large.
However the selection effects for the two types are 
usually different, and different methods 
for constructing control samples are adopted in earlier studies.
These make it difficult to estimate the impact of including Type Is. 
Moreover, the difference between the two types of AGNs 
indicates that they may be triggered by different processes \citep[e.g.][]{Jiang2016}. 
Including type Is may, therefore, mix various effects, making it
more difficult to understand the underlying processes.
Given all of these, we believe that it is better to investigate the 
two populations separately.

We do not consider satellite AGNs, because 
satellites complicate the interpretations of the results. 
To demonstrate this, we show in the right panels of 
Figure \ref{fig_2pcf} the 2PCCFs of the total AGN sample (tAGN),
in comparison to the corresponding control galaxy 
sample, tG$^{\rm c}$. As one can see, the correlation for 
tAGNs is weaker than that of the control sample on both small and 
large scales. At scales of hundreds of $\kpc$, there is a dip in 
the ratio between tAGN and tG$^{\rm c}$ \citep[see e.g.][]{Li06b}. 
Comparing to the results for centrals (green lines),
we can see that including satellites enhances the difference 
between AGNs and normal galaxies on both small and large scales. 
One possible explanation is that the AGN fraction is 
lower among satellites than among centrals \cite[e.g.][]{Li06b,WangL-19}.
Clearly, more effects have to be taken into 
account in order to explain results obtained from a mixture 
of centrals and satellites.

\subsection{The importance of using well-defined control samples}\label{sec_ipcs}

Two galaxy properties are commonly
adopted in the construction of control samples, 
one is galaxy stellar mass and the other is redshift.
For investigations using flux-limited samples to measure the 2PCCF, 
as is carried out here, it is essential to match samples to be compared  
in redshift. 
Since the galaxy population covers a large range 
of stellar mass, and since many other properties 
are related to galaxy mass, controlling galaxy 
mass is necessary if one wants to separate effects 
caused by other properties from those caused by 
the mass. Other galaxy properties, such as color, $D_n4000$ and 
$\sigma_*$, are sometimes also used to control AGN and normal 
galaxy samples, to find differences that are not caused by 
these properties. 
However, inappropriate control samples can reduce the effects 
one is looking for. For example, if galaxy interaction can 
significantly affect the bulges, as indicated by the results in 
Section \ref{sec_vd}, controlling $\sigma_*$ may lead to 
an underestimate of the role of the interaction. 
Similarly, if AGNs can strongly affect star formation 
in their hosts, comparing AGNs with normal galaxies 
that have similar color and $D_n4000$ to AGN host 
galaxies might lead to biased results.

As shown in Section \ref{sec_fq}, our results suggest that the 
processes linked to AGNs may change the properties of their host galaxies. 
If we want to investigate whether or not galaxy environment 
has played an important role in these processes, galaxies in the 
control sample should be statistically similar to the progenitors 
of the AGN hosts (i.e. the host galaxies before the onset of the AGN) 
rather than the host galaxies on the AGN duty cycle. 
This suggests that we should select, as our comparison 
sample, normal galaxies that have the same properties 
as the AGN host galaxies before the onset of AGN.   
This is not straightforward to do but our results provide some hints.
Since AGN hosts and star-forming galaxies share the same SHMR, 
controlling stellar mass is equivalent to controlling halo mass. 
Since the average formation history of central galaxies 
is determined by the host halo mass, star-forming galaxies controlled 
in stellar mass thus provide a comparison sample that we need to 
investigate whether or not the host galaxies of AGNs are special 
in their environment and evolutionary stages relative to the average 
population of galaxies. According to $N$-body simulations, halos of 
$10^{12}\msun$ at $z=0$ assembled half of their mass at $z\sim1$ 
\citep{WangH-11}, and so the growth timescale for a 
$10^{12}\msun$ halo is typically about 7 Gyrs.
This time scale is much longer than the time scale 
for the evolution between star-forming galaxies and AGNs, 
which is about 1-2 Gyrs according to the values of 
$D_n4000$ of AGN hosts and star-forming galaxies. This 
indicates that controlling halo mass also provides a 
stable reference to investigate evolution in star formation and 
AGN activities.  

Since quiescent galaxies reside in more massive 
halos than AGN hosts of the same stellar mass, 
it is not appropriate to use them to form a 
comparison sample to investigate the host galaxies of 
AGNs in their environment and evolutionary stages relative 
to the average population. Note that only a small 
fraction, about $3\sim5$ percents of galaxies with  
large $D_n4000$, at which quenched galaxies dominate, 
have strong AGN activities (Figure \ref{fig_AGNfrc}).
Assuming that quiescent galaxies are quenched long time ago,
\cite{Man19} also suggested the exclusion of quiescent galaxies in 
comparing AGNs with normal galaxies. Our results provide 
further justifications for such an approach. 

Finally, we note that our control samples contain AGNs.
In order to check whether the AGN contamination in the 
control samples can cause bias in our results, we repeat all of 
the analyses shown above by using control samples that exclude 
AGNs. Our tests show that the results change very little. 
This is expected, because only a small fraction of galaxies, 
between 5\% and 15\% (see Section \ref{sec_galpro}),  
are identified as AGNs. For simplicity, we do not show them here.

\subsection{Triggering AGN with minor interactions}\label{sec_trigger}

Comparing with star-forming galaxies, we find that local 
environment plays a dramatic role in triggering AGN activity.
In particular, we find that small
satellites dominate the environmental difference 
between AGNs and star-forming galaxies.
This suggests that minor interactions may be 
responsible for driving gas to flow into the 
galaxy center and trigger AGN activity. This does not mean 
that massive satellites can not trigger AGN activity,
but that they are not dominating because they are 
rarer. 

\begin{figure}
    \centering
    \includegraphics[scale=0.28]{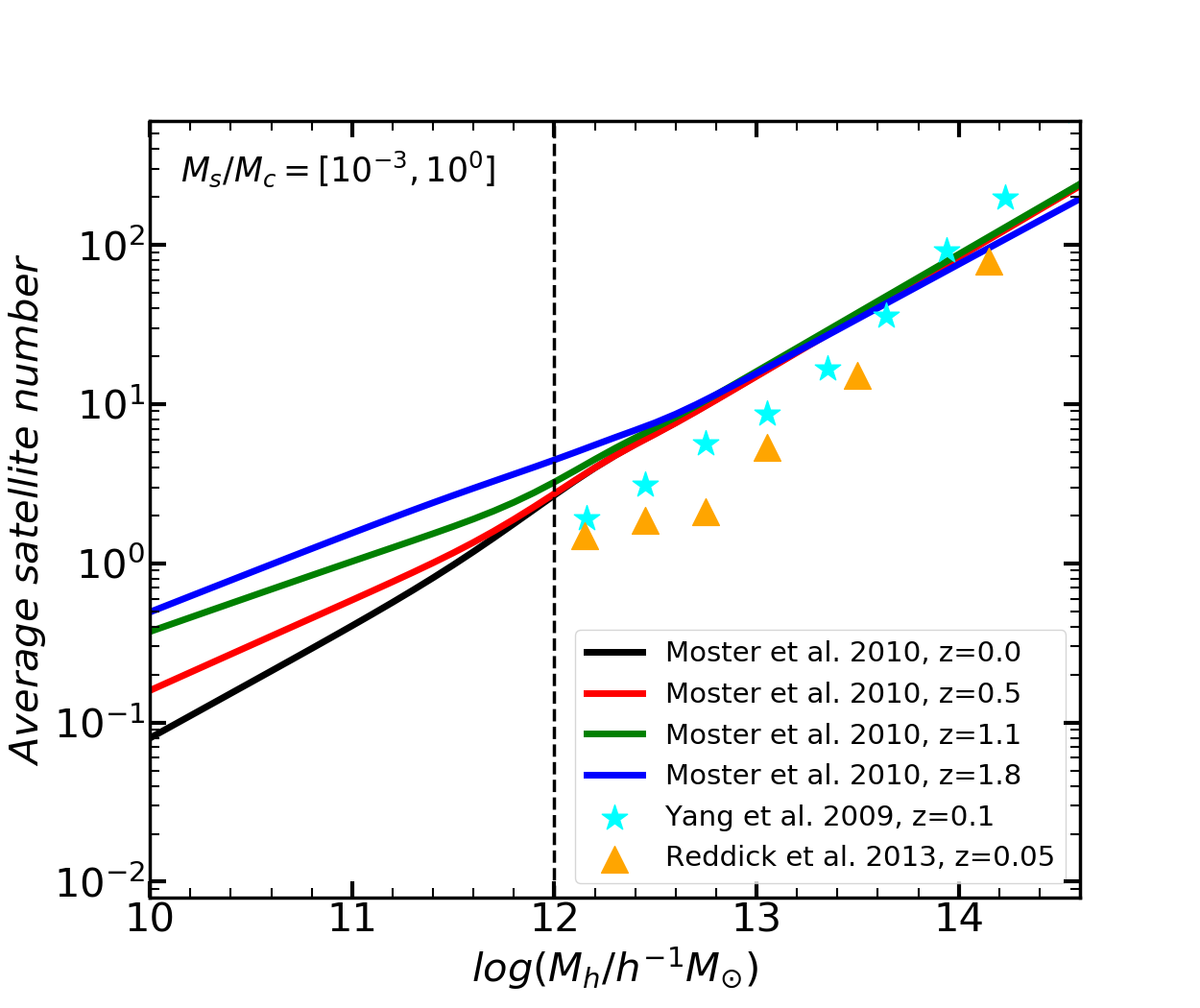}
    \caption{The average number of satellite galaxies, with $M_{\rm s}$ in range of [$10^{-3}$, 1] of $M_{\rm c}$,  as function of halo mass. Here, $M_{\rm s}$ and $M_{\rm c}$ are the masses of satellites and centrals. Four lines denote the results at four redshifts derived from the conditional stellar mass function (CSMF) shown in \cite{Moster10}. The triangles and stars represent the results derived from the CSMFs of \cite{Reddick13} and \cite{Yang-Mo-vandenBosch-09}, respectively.}
    \label{fig_satellitenumber}
\end{figure}

The probability for a central galaxy to interact with its 
satellites depends on the number 
of satellites within its host halo. Based on the 
conditional stellar mass function (CSMF) for 
satellites derived using group catalogs and 
the abundance matching methods\citep{Yang-Mo-vandenBosch-09, Moster10, Reddick13}, 
we estimate the number of satellites with 
stellar masses, $M_{\rm s}$, 
in the range of $[10^{-3}, 1]\times M_{\rm c}$,
where $M_{\rm c}$ is the stellar mass of the central galaxy,
and show the results in Figure \ref{fig_satellitenumber}. 
The mean number of satellites per halo at $\log{M_{\rm h}/\msun}=12$ is of
the order of unity at $z\sim0$, and the variance among 
different methods is not too large. 
This means that central galaxies in these halos 
have a high probability, close to unity, 
to interact with their satellites with mass 
above $M_{\rm c}/1000$ within an time scale, $t_{\rm it}$.
Since interaction requires that centrals and satellites 
are close enough, $t_{\rm it}$ is expected to be comparable 
to the merger timescale of galaxies. As shown in \cite{Jiang2008},
merger time scale ranges from 1 Gyr to about 10 Gyrs, 
with a typical value of 3 Gyrs. 

The number of satellites increases with halo mass. 
Thus, interaction is expected to be even more frequent 
for halos of mass larger than $10^{12}\msun$.
However, in these halos, most of the central galaxies are quiescent  
galaxies, containing little cold gas, 
so such interactions may not produce AGNs.   
This, together with the fact the abundance of halos 
decreases with halo mass, implies that  only a small 
fraction of optical AGNs reside in massive halos. 
For halos with $M_{\rm h}\ll10^{12}\msun$, the mean number of 
satellites is much smaller than unity, so that the 
interaction probability is also small. This
suggests that the AGN fraction among low-mass 
galaxies should also be lower, even though they
have plenty of cold gas and strong star formation. 
Figure \ref{fig_AGN_frac_sm} shows the AGN fraction as a function
of halo mass, which is converted from AGN fraction as a function of
stellar mass based on the SHMR in \cite{Yang-Mo-vandenBosch-09}.
We can see that AGN fraction reaches maximum 
around $10^{12}\msun$ and becomes
much lower at the low and high halo mass end, consistent with our analysis.
Note that the satellite number and $t_{\rm it}$ may vary 
significantly among individual halos of a 
given $M_{\rm h}$, so that the halos within which 
significant interaction happens may have a broad 
mass distribution. The number of satellites in a halo, 
which increases with halo mass, the cold gas reservoir, and the 
host halo abundance, which decreases with 
halo mass, working together, may thus make halos 
with  $\log{M_{\rm h, A}/\msun}\sim 12$ the 
most favorable places for AGN activities. 

In Figure \ref{fig_satellitenumber}, 
we also show the satellite number at different redshifts based on
the CSMF of \cite{Moster10}. The satellite 
number for $\log{M_{\rm h, A}/\msun}=12$ 
is close to unity for different redshifts,  
consistent with the weak redshift dependence of 
the host halo mass of AGNs \citep{Croom05}. 
Since the interaction time scale is expected to be proportional to halo dynamical time, 
$t_{\rm it}$ is expected to decrease with increasing redshift. Moreover, galaxies at high redshift 
are expected to contain more cold gas 
than the low-$z$ counterparts. 
These two factors together may lead to much 
stronger and more frequent AGN activities, 
which may be relevant to quasars observed at high redshift.

One question with this scenario is whether or not 
small satellites with masses down to $M_{\rm c}/1000$ 
are capable of triggering AGN activities. 
Satellites are usually surrounded by their own dark matter halos (subhalos). For host 
halos of $\log{M_{\rm h}/\msun}=12$, $M_{\rm c}\sim 10^{10.3}\msunt$ according to the SHMR,
and so $M_{\rm c}/1000\sim 10^{7.3}\msunt$. 
According to the SHMR for central galaxies, these galaxies
reside in halos of $\log{M_{\rm h}/\msun}\sim10.3$ before becoming satellites. 
Therefore, halos associated 
with these satellites may be massive enough to 
disrupt the interstellar medium in centrals and 
to induce galactic-scale gas inflow. Since 
dark halos are more extended than galaxies, 
they can be severely stripped before interacting 
with the centrals, the exact mass relevant to 
the interaction is unclear. 

It is also unclear whether or not galactic-scale 
inflows produced by the interaction with satellite 
galaxies can directly fuel the central super-massive 
black holes (SMBHs) to produce AGNs. One 
possibility is that the galactic-scale 
inflow can enhance the star formation in a galaxy 
center to build up a pseudo-bulge/bar component,
as indicated by Figure \ref{fig_2pcfvdr}, which 
in turn can halp drive cold gas toward the SMBH. 
Consistent with this, a large fraction of the host galaxies 
of AGNs are observed to contain pseudo-bulge structures \citep[e.g.][]{Bennert2015}.
Alternatively, the interactions with satellites
may distort galaxy disks, and the gravitational 
torque of the non-axisymmetric disks can help to transport cold gas into the galactic center to feed 
the SMBH \citep{Hopkins2011}. These secular structures may persist long after the original interaction, 
providing a long lasting engine to support AGN
activities. Clearly, further investigations 
are needed to verify such a scenario.

\subsection{Halo growth, Interactions, AGN feedback and Galaxy evolution}\label{sec_fq}

\begin{figure*}
    \centering
    \includegraphics[scale=0.67]{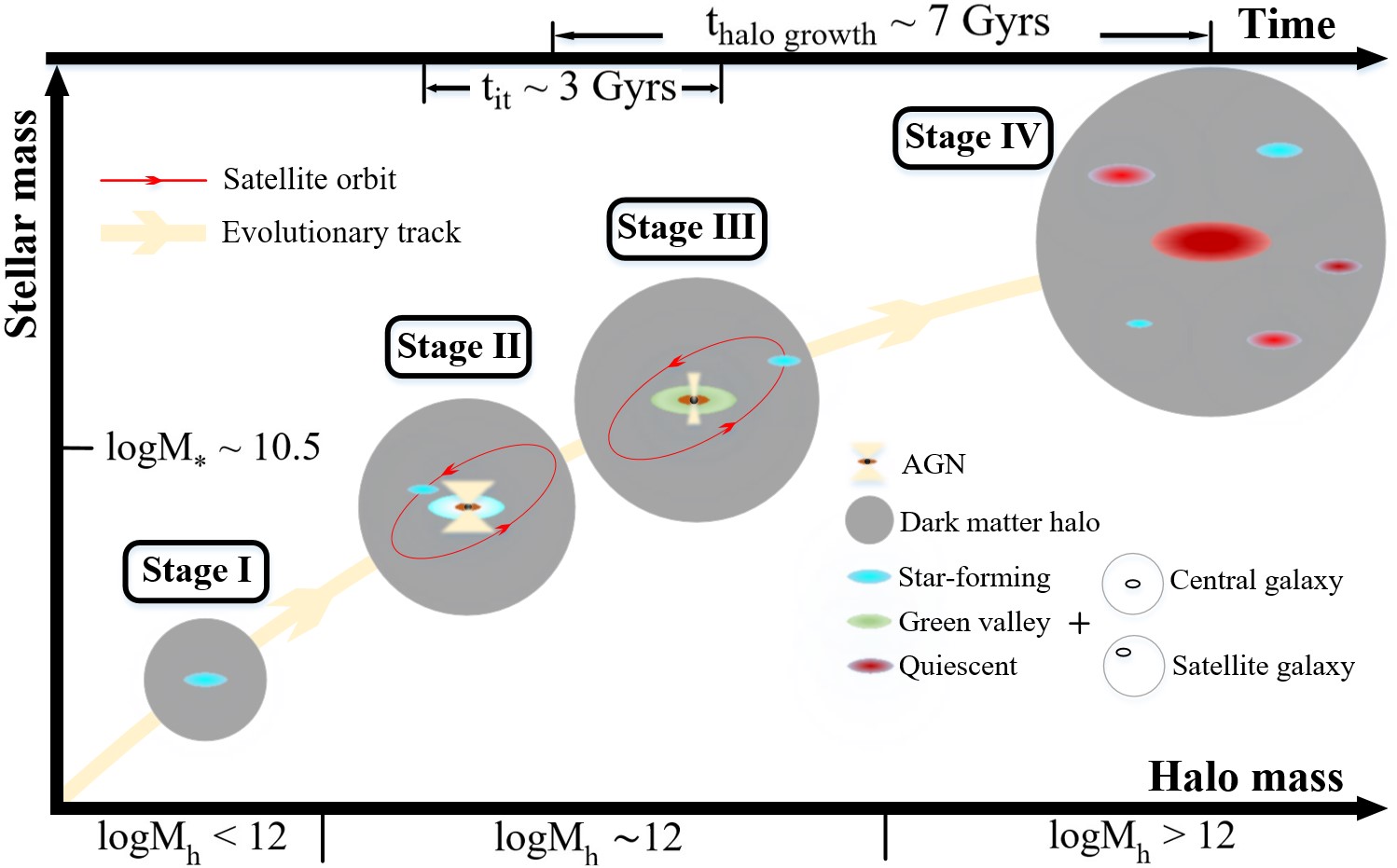}
    \caption{Sketch of galaxy-halo evolution and AGN triggering through minor interactions. At stage I when halo mass is much less than 10$^{12}\msun$, the halo may only host one central galaxy which contains plenty of cold gas. No significant AGN activity is detected because of the absence of the minor interaction with satellites. At stage II when the halo mass grows to $\sim$10$^{12}\msun$, the number of satellites per halo reaches the order of unity so that interaction between central and satellites may happen within a time scale $\sim$3 Gyrs, and consequently makes the cold gas in the disk flow into the galaxy center, triggers AGN activities and strong star formation (or even a star burst). At stage III, star formation in galaxy center has been quenched by star burst and AGN activity and the stellar population formed in the star burst evolves from $D_n4000=1$ to 1.5. The non-axisymmetric structures produced by interaction may continue to drive cold gas inwards, albeit at a reduced rate, producing a low-level AGN multiple times. Since the amount of cold gas is reduced by the star burst, no significant enhancement of star formation is expected during this stage. At stage IV, after a typical timescale of 7 Gyrs, the host halo mass becomes much lager than 10$^{12}\msun$, the star formation in central galaxy is fully quenched owning to the lack of cold gas, and AGN may still be triggered by the secular process but too weak to detect.}
    \label{fig_evolution}
\end{figure*}

The efficiency for converting baryonic gas into stars, 
which can be characterized by $M_{\rm c}/M_{\rm h}$, 
peaks at a mass $\log (M_{\rm h, p}/\msun)\sim 12$. 
\cite{WangE-18a} found that the quenched fraction for centrals 
increases with $M_{\rm h}$ very quickly around $M_{\rm h, p}$.
These results imply that a large 
fraction of central galaxies have their star formation 
quenched when their halos reach $M_{\rm h, p}$ 
\citep[see more discussion on this threshold mass in][]{Dekel-Birnboim-06, Gabor-Dave-15}. 
Interestingly, halos of $10^{12}\msun$ are also important for AGN activities, as shown in Section \ref{sec_hmng}
\citep[see also][]{Croom05, Mandelbaum09}. 
One scenario proposed in the literature
is that AGN feedback can quench the star formation 
in their host galaxies. This is supported by the 
high AGN fraction at $D_n4000=1.5\sim1.6$ 
(Figure \ref{fig_AGNfrc}), where the transition 
from star-forming to quiescent population occurs. 
Since the timescales for AGN activities and 
quenching are expected to be shorter
than that for halo growth, one would expect 
that AGNs have the same SHMR as star-forming galaxies, 
consistent with our results in Figure \ref{fig_shmr}.
However, there are unresolved problems in 
this scenario. In the local Universe, AGN radiation 
and their winds are usually weak \citep{Kauffmann-09}, 
and it may be difficult to expel cold gas from host galaxies. 
Moreover, the timescale for galaxy 
quenching is about 1 Gyr, while the timescale for 
one cycle of AGN activity is much smaller.
Thus, the observed AGNs are unlikely the ones 
that expelled the cold gas and quenched the star formation.

Based on our results, we propose a scenario
which is sketched in Figure \ref{fig_evolution}.
In this scenario, central galaxies in halos of $\log(M_{\rm h}/\msun)<12$ 
are mostly star-forming galaxies \citep[see e.g.][]{WangE-18a}, 
and thus contain large amounts of cold gas. 
Since the number of satellites per halo 
in these halos is small, most of the galaxies 
exhibit no significant AGN activity. 
When halos grow to about $10^{12}\msun$, the number of 
satellites reaches the order of unity, and close interaction 
happens within a time scale  $t_{\rm it}\sim 3$ Gyrs. Interaction makes the 
cold gas in the disk flow into the galaxy center, 
triggering AGN activities and strong star 
formation (or even a star burst).
The AGN associated with the star burst usually 
has a high accretion rate and luminosity \citep[][]{Kauffmann-09, Greene2020}, 
and thus can launch strong winds \citep{WangH-AGN-11}
that can quench the star formation in galaxy center and shut off 
the fuel supply to the AGN. Because of this, 
the stellar population formed in the star burst 
evolves from $D_n4000=1$ to 1.5 within about 
1 Gyr. During this period, secular non-axisymmetric  
structures produced by the interaction
may continue to send cold gas from the disk, 
albeit at a reduced rate, to feed the SMBH, 
producing a low-level AGN. In this case, low 
level AGNs may be triggered multiple times 
by the secular evolution, so that the total 
duty cycle time is much longer than that of 
a single cycle. Since the total amount of cold 
gas is already reduced by the star burst,  
the AGNs triggered by the secular evolution  
are expected to be weak, and no significant 
enhancement of star formation is expected
from the process. As the process proceeds, 
the host galaxy will become poorer in cold gas,  
eventually becoming quenched in star formation, 
and the AGN triggered by the secular structure 
will become too weak to detect. During this 
long time scale, halos can still grow 
significantly while galaxy mass grows little, 
which may explain why quiescent galaxies
have more massive halos than star-forming galaxies 
of the same stellar mass.

In this scenario, most of the AGNs observed in 
the local Universe are not associated with star bursts, 
although their host galaxies may have gone 
through star burst phases earlier. Indeed, the 
analyses on the H$\delta$ absorption line 
have revealed that a significant fraction of AGNs 
reside in post-star-burst galaxies \citep{Kauffmann-AGN-03}.
The early star burst in the center of a galaxy can 
help to build up a central bulge. 
This may explain why galaxies of larger 
$\sigma_*$ on average are surrounded by 
larger number of satellites (Figure \ref{fig_2pcfvdr}). 
The connection between star burst and AGN activity is also 
supported by the high AGN fraction in galaxies with $D_n4000\sim 1$ 
(Figure \ref{fig_AGNfrc}). Since AGNs in star burst galaxies ($D_n4000<1.2$) 
are directly triggered by galaxy interaction, while AGNs hosted by 
galaxies in the transition from star-forming to quiescent populations 
($D_n4000\sim1.5$) are mainly driven by secular evolution, 
two characteristic time scales are relevant here: the star 
burst time scale, which is typically $10^8$ years, and the secular 
evolution time scale, which is typically a few Gyrs for present-day 
galaxies. The two peak distribution shown in Figure \ref{fig_AGNfrc}
may be explained by these two time scales.

\section{Summary}\label{sec_sum}

Based on spectroscopic and shear data of SDSS galaxies 
in local Universe, we investigate the difference and 
similarity between optically-selected AGNs
and normal galaxies. Here we only focus on the central 
galaxies of galaxy groups and clusters. We construct 
control samples for AGNs from quiescent and 
star-forming galaxies, respectively, so that we can 
inspect the location of AGNs in the evolutionary 
path of galaxies. We investigate the galaxy properties, 
such as star formation rate, color, $D_n4000$ and 
central velocity dispersion, 
for AGNs and control samples. We use cross-correlation 
and weak lensing measurements 
to constrain the halo masses and surrounding 
satellites of these galaxies.
Our main scientific results are summarized as follows.

\begin{itemize}
    \item The color and $D_n4000$ distributions for the majority of AGNs are almost independent of the stellar mass (Figure \ref{fig_galpro}). In contrast, star-forming and quiescent galaxies exhibit strong or significant dependence. AGNs have larger (smaller) central velocity dispersion than star-forming (quiescent) galaxies.
    \item There are two peaks in the distribution of AGN fraction(Figure \ref{fig_AGNfrc}). One peak is at $D_n4000\sim1$ and the other at $D_n4000\sim1.5$. AGN fraction at the first peak strongly depends on stellar mass, ranging from 5\% to $>$40\%, while that at the second peak is around 30\%, almost independent of stellar mass.
    \item Combining cross-correlation function and weak lensing signal together, we measure the host halo masses for AGNs, control star-forming and quiescent galaxies. This technique significantly increases the signal-to-noise ratio for halo mass measurement(Table \ref{tab_01}).
    \item The mean host halo mass for AGNs is around $\log(M_{\rm h, A}/\msun)=12$(Table \ref{tab_01}). It is similar to the pivot halo mass in the stellar mass-halo mass relation. 
    \item AGNs and control star-forming galaxies share the same stellar mass-halo mass relation, while quiescent galaxies reside in more massive host halos than the other two populations (Figure \ref{fig_shmr}). 
    \item AGNs are surrounded by more satellites than star-forming galaxies of the same stellar mass(Figure \ref{fig_2pcf} and \ref{fig_st}). And the difference is dominated by small satellites with masses down to $10^{-3}$ of the central stellar mass(Figure \ref{fig_msmc}).
    \item For galaxies with mass similar to the host galaxies of AGNs, galaxies with larger central velocity dispersion are surrounded by more satellites (Figure \ref{fig_2pcfvdr}). 
    \item Control samples have significant impact on the environmental study for AGNs(Figure \ref{fig_2pcf}). It is required to take into account the evolutionary stage when constructing control samples.
\end{itemize}

AGN activities, galaxy quenching and the change of environment (halo mass and satellites)
occur at different time scales, ranging from $\ll 0.1$ Gyr to several Gyr. 
Our results clearly show that, on the timescale for galaxy quenching, 
optical AGNs look different from normal galaxies. AGNs like to 
reside in star burst galaxies and
`green valley' galaxies that are transiting from star-forming to quiescent phase.
However, on a long timescale for halo growth, AGNs are close to 
star-forming galaxies, but very far from quiescent galaxies in the evolutionary path of galaxies. If the timescale for AGN activities 
is really very short, less than 0.1 Gyr, as claimed in the literature,
multiple AGN activities for one single SMBH are required to explain 
the high AGN fraction at some specific evolution stages.

We thus propose a scenario, in which minor interactions with small satellites and
their dark halos, as well as the warped and unstable galactic structures 
caused by the interactions, can trigger gas inflow to ignite the AGNs multiple times. 
The first interaction with satellites can cause strong star formation,
even star burst, in galactic center, which may help to buildup the bugles, 
and trigger AGNs of high luminosity. The feedback from the strong AGNs and star burst 
ceases the star formation. It can explain various observational 
facts shown in this paper and previous studies. 
Interaction probability is dependent on satellite number. 
We find that the mean number of satellites with 
$M_{\rm s}>M_{\rm c}/1000$ strongly increases with 
halo mass and reaches about unity around halo mass of $10^{12}\msun$. 
Together with the star-forming population and halo abundance quickly 
declining with halo mass, our scenario provides a natural explanation 
on why optically selected AGNs favor halos of $10^{12}\msun$.
It may also help to yield the pivot mass in the stellar mass-halo mass relation.

Besides the minor interaction scenario, other processes may 
also work. For example, major mergers and violent interactions 
can induce strong gas perturbation and trigger AGN activity. 
Internal secular evolution forms galactic bars, which cause 
gas inflow to feed the SMBH. The elliptical shape of halos, 
which is also correlated with substructure number, can also lead to a 
non-axisymmetric gravitational potential and cause slow gas inflow. 
Moreover, the release of the gravitational energy of these satellites
may also help to maintain the circumgalactic gas hot and tenuous. 
Detailed studies are certainly required to understand the role 
and efficiency of various mechanisms.

\section*{Acknowledgments}
We thank the referee for useful comments. This work is supported by the National Key R\&D Program of China (grant No. 2018YFA0404503), the National Natural Science Foundation of China (NSFC, Nos.  11733004, 11421303, 11890693, 11522324, 11773032 and 12022306), the National Basic Research Program of China (973 Program)(2015CB857002), and the Fundamental Research Funds for the Central Universities. The work is also supported by the Supercomputer Center of University of Science and Technology of China.

\bibliographystyle{aa}
\bibliography{rewritebib.bib}

\end{document}